\documentclass[%
 reprint, superscriptaddress, 
 amsmath,amssymb, aps, prmaterials, 
 ]{revtex4-2}
\usepackage{graphicx}
\usepackage{multirow}
\usepackage{comment}
\usepackage[usenames]{color}
\usepackage{graphicx}
\usepackage{dcolumn}
\usepackage{bm}
\usepackage[colorlinks=true, allcolors=blue]{hyperref}

\newcommand{\br}{{\bm r}}
\newcommand{\sigmaY}{{\sigma_{\tt Y}}}

\begin{document}

\preprint{APS/123-QED}

\title{Temperature dependence of fast relaxation processes in amorphous materials 
}
\author{Gieberth Rodriguez-Lopez}
\affiliation{Instituto de Nanociencia y Nanotecnolog\'{\i}a, CNEA--CONICET, 
Centro At\'omico Bariloche, R8402AGP S. C. de Bariloche, R\'{\i}o Negro, Argentina.}
 \email{gieberth.rodriguez@cab.cnea.gov.ar}
 
\author{Kirsten Martens}%
\affiliation{Univ. Grenoble Alpes, CNRS, LIPhy, 38000 Grenoble, France.}

\author{Ezequiel E. Ferrero}%
\affiliation{Instituto de Nanociencia y Nanotecnolog\'{\i}a, CNEA--CONICET, 
Centro At\'omico Bariloche, R8402AGP S. C. de Bariloche, R\'{\i}o Negro, Argentina.}
\affiliation{Departament de Física de la Matèria Condensada, Universitat de Barcelona, Martí i Franquès 1, 08028 Barcelona, Spain.}
\affiliation{Institute of Complex Systems (UBICS), Universitat de Barcelona, Barcelona, Spain
}
\email{ezequiel.ferrero@ub.edu}
 

\begin{abstract}

We examine the structural relaxation of glassy materials at finite 
temperatures, considering the effect of activated rearrangements and 
long-range elastic interactions. 
Our three-dimensional mesoscopic relaxation model shows how the displacements
induced by localized relaxation events can result in faster-than-exponential 
relaxation.
Thermal activation allows for local rearrangements, which generate
elastic responses and possibly cascades of new relaxation events. 
To study the interplay between this elastically-dominated and thermally-dominated 
dynamics, we introduce tracer particles that follow the displacement field induced by the local relaxation events and also incorporate Brownian motion. 
Our results reveal that the dynamic exponents and shape parameter of the dynamical
structure factor depend on this competition and display a crossover from 
faster-than-exponential to exponential relaxation as temperature increases, 
consistent with recent observations in metallic glasses. 
Additionally, we find the distribution of waiting times between activations
to be broadly distributed at low temperatures, providing a measure of dynamical 
heterogeneities characteristic for to glassy dynamics.

\end{abstract}

\maketitle

\section{Introduction}

When we rapidly cool down a metallic or polymeric melt to temperatures 
below the glass transition, the result is a highly viscous, 
heterogeneous and frustrated material that we call `amorphous solid'.
Its non-equilibrium, topological and dynamical properties
are part of one of the most salient open problems in 
Statistical Mechanics and, also, in Materials Science.
From the viewpoint of the microscopic structure, some regions of
the material freeze in states with high local stress barriers and 
others are more easily prone to relax since they correspond to soft 
regions close to instability.
Although the material appears now solid on long time-scales and responds 
elastically to small deformations, it may still exhibit measurable internal 
relaxation dynamics. 
The relaxation process that develops and eventually alters the glass physical
properties involves a wide range of time, energy and length scales, 
aging processes and sample preparation dependencies.
Understanding this spontaneous aging is of key relevance in the attempts 
to control such mechanical degradation; for instance, to manage these materials 
in industrial applications.
This makes it a relevant problem not only from a theoretical but also from 
a practical point of view.

The complexity of a quiescent glass relaxation is usually quantified by 
the way in which the so-called ``dynamic structure factor'' 
(\textit{viz.} the intermediate scattering
function) deviates from an exponential relaxation. 
In many cases, {\it stretched exponentials} are observed, possibly resulting from a broad 
distribution of relaxation times due to material heterogeneity. 
However, the opposite situation of `compressed' relaxation 
(faster than the exponential) is also observed experimentally 
in various materials, such as metallic glasses or colloidal gels. 

These {\it compressed exponentials} have recently been interpreted within a 
framework that puts forward the elastic response to local relaxation events in the material.
The suggestion is that, even when the bulk material has an intrinsic elastic 
nature, the occurrence of localized and rather sparse relaxation events can be 
enough to modify and dominate the global relaxation behavior of the system.
Each event leads to an elastic reponse of the sourrounding material and induces thereby a long range displacement field for the sourrounding particles.
An experimental framework for the study of this unusual relaxation 
was first established by dynamic light scattering (DLS) works in 
fractal colloidal gels~\cite{CipellettiMaBaWe-PRL2000, RamosCi2001, CipellettiRaMaPiWePaJo-FD2003, CipellettiRa-JPCM2005, DuriCi-EPL2006}.
More recently, x-ray photon correlation spectroscopy (XPCS) has been used 
to study slow dynamics following the same spirit, but this time not only 
`soft' materials as colloidal 
suspensions~\cite{CaronnaChMaCu-PRL2008, DuriAuStLeChGrGu-PRL2009}, 
colloidal glasses~\cite{AngeliniZuFlMaRuRu-SM2013} and 
gels~\cite{TrappePiRaRoBiCi-PRE2007, OrsiCrBaMa-PRL2012}, but also in 
hard-amorphous materials such as 
metallic glasses  (MGs)~\cite{RutaChMoCiPiBrGiGo-PRL2012, ruta2013, evenson2015}.
An overview of the state of the art by 2017 in physical aging and relaxation
processes in MGs can be found in reference~\cite{RutaJPCM2017}.

The structure factor in these experiments is extracted from the time-averaged 
temporal auto-correlation function of the scattered intensity 
$g^{(2)}(q,t)=\left<I(q,\tau)I(q,\tau+t)\right>/\left<I(q)\right>^2$
where $I(q,t)$ is the intensity of the signal measured with the 
corresponding scattering technique.
$f(q,t)$ is related with $g^{(2)}$ through the Siegert relation
\begin{equation}
  g^{(2)}(q,t) - 1 = \kappa | f(q,t) |^2 \; .
\end{equation}
The common feature observed in several experiments is a structure factor 
showing a compressed exponential behavior in time $t$ and 
scattering vector $q$~\cite{CipellettiRa-JPCM2005,RutaJPCM2017}:
\begin{equation}
 f(q,t) \sim \exp[-(t/\tau_f)^\beta]\;, 
\end{equation} 
 
 \noindent with $\tau_f \sim q^{-n}$, $n \sim 1$ and $\beta>1$\;.
The observed dynamics was \textit{a priori} unexpected, not only because of such a
faster than exponential decay of the correlations but also because it contrasts 
with the usual diffusive behavior at long time-scales ($\tau_f \sim q^{-2}$) found in molecular 
dynamics simulations of most glassy systems.
It should be mentioned that, although this measurements are performed in 
aging system, the compressed relaxation is not intrinsically an 
non-equilibrium feature.
In general, experiments are focused in a time-window  where the waiting 
time dependence can be disregarded, the age 
of the material can be considered fairly unchanged during the measurement.
Moreover, cyclic shear experiments~\cite{Tamborini-PRL2014} showed that these 
compressed-exponential relaxation dynamics associated with ballistic-like motion is observed even in a well 
controlled stationary state.
Note, that even in these examples of stationary dynamics a dynamical competition of time-scales remains present, that we show to be a possible origin of compressed exponential relaxation.
In the stationary setup (without any applied deformation) thermal agitation causes Brownian motion of the individual particles within their cages and eventually local relaxation events are triggered by thermal activation. 
The elastic response to these events can induce further relaxation events, by destabilising new regions, that were already close to instability.
Higher temperatures, not only induce more and more frequent relaxation events but also 
increase the diffusion of the particles.
As we will discuss, the interplay between thermal Brownian motion and the persistent motion induced by elastic
displacement fields due to localised relaxation events affects and quantitatively determines the relaxation process. In particular, this competition 
leads to an {\it effectively temperature dependent exponent} $\beta$ in the compressed 
exponential behavior of the dynamical structure factor $f(q,t) \sim \exp[-(t/\tau_f)^\beta]$.

\subsection{A literature overview}

The idea that relates compressed exponential behavior of $f(q,t)$ with the elastic reponse to localized relaxation events 
events was introduced by Cipelletti et al. in~\cite{CipellettiMaBaWe-PRL2000}.
It was later reformalized by Bouchaud \& Pitard within in a mean-field 
scenario~\cite{BouchaudPi-EPJE2001, BouchaudPi-EPJE2002, Bouchaud-InBook2008}.
In~\cite{FerreroPRL2014} part of the authors of the present manuscript have proven this scenario valid in finite 
dimensions (2D) within simulations of a mesoscopic model for the relaxation dynamics of glassy materials.
Although the local rearrangements leading to relaxation of the material are in 
general not individually assessed in the experiments, their 
relation with the compressed exponential behavior has been widely acknowledged.

Luo et al.~\cite{LuoPRL2017} explored a wide temporal and temperature range in the 
relaxation of three typical Zr- and La-based metallic glasses (MGs). 
They directly measure the stress by applying a small tensile
deformation on MG ribbons and follow its relaxation in time.
They find a gradual change of the relaxation profile from a single-step to a 
two-step decay upon cooling. 
They relate the first faster relaxation process to the anomalous stress-dominated
microscopic dynamics, and the secondary slower one, to subdiffusive motion at larger 
scales with a broader distribution of relaxation times.
For this stress relaxation, they also observe compressed exponentials as soon
as $T < 0.9 T_g$, where $T_{g}$ is the glass transition temperature.

In~\cite{AminiPRM2021} Amini et al. studied structural relaxation a bulk metallic glass 
forming alloy. 
Upon heating across the glass transition, the intermediate scattering function (ISF) changes from a compressed to a stretched decay, with a smooth variation of the stretching exponent and the characteristic relaxation time.
Authors relate this to a progressive transition between a stress-dominated dynamics of 
glasses and the mixed diffusion and hopping particle motion of super-cooled alloys.
In this study compressed exponentials 
are observed bellow $T_g$, in agreement with ~\cite{RutaChMoCiPiBrGiGo-PRL2012}. 
This is when the metallic alloy responds as an amorphous solid and localized
relaxation events, akin to shear transformation zones (STZs) in sheared amorphous solids, can produce persistent displacements in their surroundings. 

Interestingly, a very recent work by Song et al.~\cite{SongPNAS2022} 
investigates again the kind of soft matter systems where compressed 
exponential relaxation were first reported.
By analyzing microscopic fluctuations inside an arrested gel they differentiate among two distinct relaxation mechanisms: 
quiescent relaxations governed by the buildup of internal stresses during arrest, and perturbation-induced avalanche relaxation 
events governed by mechanical deformations in the system. 
In the quiescent case, when internal stress heterogeneities generated 
during arrest are released, they cause local strain propagation.
In this gel these rearrangements caused by the pre-stressed states are even considered 
to be athermal, with an occurrence rate exceeding strongly the one of thermally activated rearrangements. 
XPCS on the arrested gel probed at quiescence shows a second-order 
correlation function $g_2(q,t)$ decaying as a compressed exponential 
function of $(q^{1.07} t)$ in a range of $q$ values, with a compressed 
exponent $\beta\sim 1.57$. 
The notion that intermittent plastic activity happens due to mechanical aging and relaxation of pre-stresses more than thermal activation, 
can be also found in metallic glasses~\cite{evenson2015}.

Still, it should be also mentioned that not always the compressed exponent 
behavior is said to be related to relaxation of inner stresses. 
In ref.~\cite{GabrielJCP2015} the authors studied a system of polystyrene spheres in 
super-cooled propanediol.
By means of multi speckle-dynamic light scattering experiments, at 
low temperatures, compressed exponential decays are observed. 
The speckle pattern shows indication for convection in the sample due to 
a slight temperature gradient across the sample cuvette mounted in a cold 
finger cryostat. 
Authors attribute the compressed exponentials to such convection, an
effect that increases with decreasing temperature. 

On the modeling side, the phenomenon of compressed exponential relaxation 
has also been discussed recently, both in the gel context as well as for hard 
amorphous materials.
In a work on a gel created as a kinetically arrested phase separation it has been 
shown that its relaxation is accompanied by super-diffusive particle motion and compressed exponential relaxation of 
time correlation functions~\cite{chaudhuri2017ultra}. 
Spatiotemporal analysis of the dynamics reveals intermittent heterogeneities
producing spatial correlations. 
Another work on gel relaxation in a network forming attractive gel showed similar 
behavior when local bond breakings were induced by hand~\cite{Gado2017}. 
The authors evidenced a cross-over of the shape exponent from compressed to 
stretched exponentials as a function of temperature pointing towards a 
competition between Brownian motion and elastic effects through the 
relaxation of internal stresses.
On the metallic glass-former materials side, simulations in~\cite{WuNC2018}
showed that the relaxation dynamics is directly related to the local arrangement 
of icosahedral structures: Isolated icosahedra give rise to a liquid-like 
stretched exponential relaxation whereas clusters of icosahedra lead to 
a compressed exponential relaxation.

Recently, Ref.~\cite{TrachenkoJPCM2021} revisited the idea of elastically 
interacting local relaxation events~\cite{CipellettiMaBaWe-PRL2000, BouchaudPi-EPJE2001} 
in an analytical atomistic approach, and discussed both slow stretched-exponential 
relaxation and fast compressed-exponential relaxation; 
the latter being related to the `avalanche-like dynamics' in the low-temperature 
glass state.
Based on Arrhenius-type activation of stress-relaxation events, similar to the work in ref.~\cite{FerreroPRL2014}, 
their model, already at the mean-field or `one-site' approach, evidences a temperature 
dependency in both the stretched and compressed behavior regimes.

\subsection{Our work}

In this work, we first confirm through the numerical study of a simple three dimensional lattice model that compressed exponential 
relaxation can result from elastic relaxations responding to thermally induced local rearrangements akin to shear transformations 
observed in yield stress materials with external driving~\cite{lemaitre2014structural, chacko2019slow, lerbinger2022relevance}.

We use a three-dimensional elasto-plastic model of amorphous solids
(described in Sec.~\ref{sec:model}) together with the construct of 
imaginary tracer particles evolving in parallel.
These particles follow the vector displacement field generated by the 
elasto-plastic model, associated with the stress response to the thermally 
induced plastic activity of the modeled material.
The particle trajectories are then used to calculate both the mean 
square displacement and the dynamical structure factor.

Our results show that for sufficiently short times there is a super-diffusive regime 
in the mean square displacement of tracer particles, after which 
we enter a crossover regime towards diffusive behavior. 
Note that we refer to short and long times here within our coarse-grained elasto-plastic description~\footnote{Our model is not constructed to resolve the details of the dynamics occurring in the core of a relaxation event. Compared to particle-based dynamics, the short time dynamics of our model are on timescales larger than the typical beta-relaxation time of a glass-former and refer rather to the cage breaking regime. The persistent motion that we observe on ``short time scales'' is distinct from the commonly observed ballistic motion at times smaller than the beta relaxation time~\cite{karmakar2016overview}.}.
The crossover is dominated by the typical duration of plastic events. 
In addition, a compressed exponential relaxation, reminiscent of 
experimental observations, is obtained in the dynamic structure factor 
associated with the super-diffusive (ballistic-like) regime. 
At long times, in the diffusive regime, the relaxation is instead exponential. 
Furthermore, we analyze in more detail the relaxation dependency on the temperature. 
The presence of finite temperature that allows for the activation of plastic events, 
also generates thermal agitation that competes with the persistent movement on 
short times~\cite{Gado2017}.
We observe that temperature modifies the crossover between super-diffusive and diffusive motion 
and effectively generates an intermediate range of values for the exponent $\beta$ of the 
structure factor decay.
Hence, our model allows for a theoretical interpretation of recent observations in metallic glasses~\cite{RutaPRL2012,AminiPRM2021}.
We discuss this phenomenology presenting results and scaling laws for the mean-square 
displacements $\left<r^2\right>$ of tracer particles, the displacement distributions $P(u)$
in different time windows, the distributions of reactivation times of local plastic activation
$\Psi(\tau_{\tt re})$ and the dynamical structure factor evolution $S(t)$.

Originally inspired in the dense phase of athermal amorphous
materials, elasto-plastic models~\cite{NicolasRMP2018} are not 
{\it a priori} expected to capture the physics of the glass 
transition. 
Yet, building on the idea that highly viscous liquids should be 
considered as ``solids which flow''~\cite{DyrePRB1996, DyreJNCS2006, DyreRMP2006},
recent endeavors have revealed that simple EPMs with thermal
activation of plastic events are able not only to reproduce compressed 
exponential relaxation~\cite{FerreroPRL2014}, but also other 
features of glassy dynamics, such as dynamical heterogeneities and 
the emergence of dynamical correlations~\cite{OzawaPRL2023}, 
precisely due to the mediation of elastic interactions in the material.
Within this context it seems thus justified to argue that the low 
temperature limit of our model mimics reasonable well the relevant 
relaxation processes in the low-$T$ phase of glass-forming liquids. 
In this picture relaxation is dominated by displacement fields rooted in elasticity and 
local rearrangements. 
At higher temperatures it is the increasing Brownian motion of particles that leads to the breakdown of the elasto-plastic picture, 
localisation and activatation, the main features of glassy glassy dynamics, become more and more irrelevant and we enter the dynamics of high-temperature super-cooled liquids.

The next Section~\ref{sec:meanfield} is a summary of the mean-field 
arguments for the elasticity-mediated compressed exponential 
phenomenon. Although important for the general understanding of the importance of elasticity in the relaxation processes, 
this part is not directly needed to access the main part of this manuscript describing our work on the spatially-resolved elasto-plastic modeling.

In Section~\ref{sec:model} we present our elasto-plastic model 
and the construction of thetracer particles for the particle displacements.
Section~\ref{sec:results} presents our findings and 
we conclude in Section~\ref{sec:discussion}.

\section{Recall of mean field arguments}
\label{sec:meanfield}

In~\cite{CipellettiMaBaWe-PRL2000} an heuristic explanation for the phenomenon
of compressed exponentials was first introduced,
based on the ``syneresis'' of a gel.
Syneresis is a spontaneous contraction of a gel, which occurs locally,
accompanied by expulsion of liquid from a pore. The gel shrinks. 
Such mechanical inhomogeneities act as local dipole forces with a long range 
elastic impact on its surroundings, creating a complex deformation field~\cite{CipellettiMaBaWe-PRL2000}.
The argument leading to the justification of a compressed exponential observation
in the dynamical structure factor goes as follows.

The displacement field $u$ due to an inhomogeneity (a syneresis event) at 
a distance $r$ goes as
\begin{equation}\label{eq:displdipole}
    u \sim \epsilon(t)V_1 / r^{d-1}
\end{equation}
where $V_1$ is an estimate for the volume of the region involved in the syneresis, 
$\epsilon(t)$ is the evolving strain on that region, and $d$ is the dimension of the system.
If the inhomogeneity is placed at the origin, the generated displacement field decays algebraically.
The observable that we are interested in measures spatial correlations at two different times, an corresponds to an overlap function.

\begin{figure}[t!]
\includegraphics[width=0.40\textwidth]{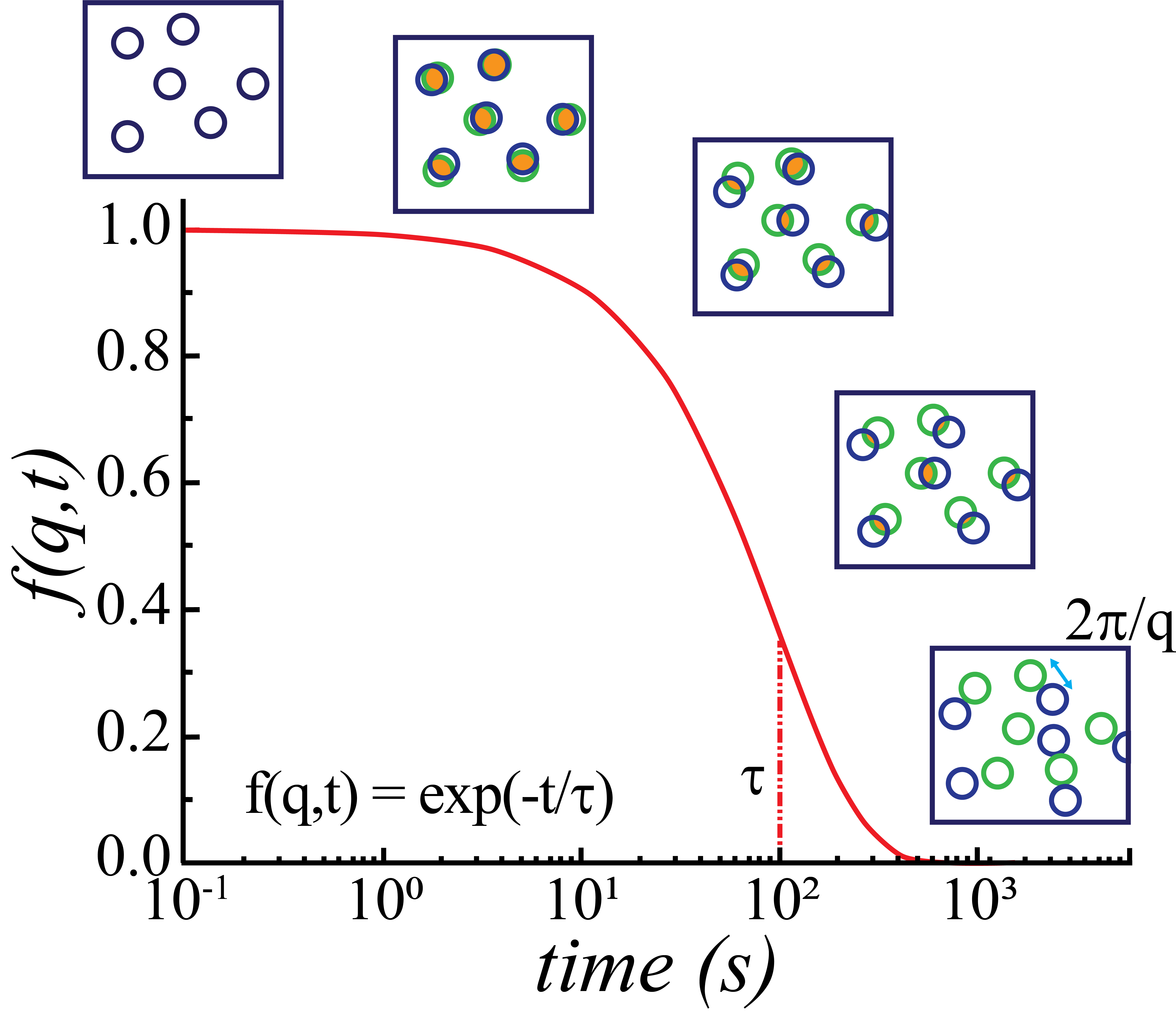}
\caption{\label{fig:schematicRuta}
Schematic figure that shows that given a spatial configuration of particles ($t=0$),
if the particles move the structure factor decays until this system decorrelates 
the initial condition in a time $t$. (adapted from a slide by B. Ruta)
}
\end{figure}

Given an initial configuration of particles, as depicted in Fig.~\ref{fig:schematicRuta}, the
structure factor $f(t,q)$ would stay close to $f(t=0,q)=1$ if the particles don't move, and will
drop all the way down to $0$ if the configurational correlation with the initial condition
is completely lost at time $t$.
Due to the intensity decay of the displacement field, the particles closer to the
inhomogeneity will move stronger and lead to a strong decrease of the
spatial correlations.
If we are observing the system at a typical space resolution of $q^{-1}$ 
(akin to using light scattering techniques with a wavevector of modulus $q$)
and we ask for $u<q^{-1}$ to consider that something has basically `not moved' 
at that length-scale,
we need it to be at a distance to the closest dipole equal or larger than
\begin{equation}
    r_{\tt min} \simeq \left[q \Delta_\epsilon(t) V_1\right]^\frac{1}{d-1}.
\end{equation}
This simply accounts to inverting eq.\ref{eq:displdipole} and introducing
$\Delta_\epsilon(t)\simeq q^{-1}\epsilon(t)$, as a measurable linear displacement.
The distance to the closest dipole will depend on the density of syneresis events.
Assuming an inhomogeneity concentration $c$ in a system volume in dimension $d$, we can expect
the probability of being further than a distance $r_{\tt min}$ to any inhomogeneity to decay as
\begin{equation}
    P(r_{\tt min}) \simeq \exp\left[{-c r_{\tt min}^d}\right] \simeq 
    \exp\left[{-c \left[q \Delta_\epsilon(t) V_1\right]^\frac{d}{d-1}}\right].
\end{equation}
Notice that, $P(r_{\tt min})\equiv P_{r_{\tt min}}(q,t)$ is already a good 
proxy for $f(q,t)$, since higher probability of being away from any strong 
field distortion means higher preservation of correlation at a given time.
So one can propose $f_{\tt mf}(q,t) \propto P_{r_{\tt min}}(q,t)$, and assuming that
the local strain within the inhomogeneity varies linearly in time $\epsilon(t) \simeq at$
we get
\begin{equation}\label{eq:fqt_compressed}
    f_{\tt mf}(q,t) \propto \exp\left[ -(t/\tau_r)^{d/(d-1)} \right]
\end{equation}
with $\tau_r = c^{-(d-1)/d} (V_1 a q)^{-1}$ emerging as a characteristic relaxation time.
Notice that $\tau_r \propto q^{-1}$, which is typical of a persitent motion.
Generalizing Eq.~\ref{eq:fqt_compressed} to three dimensions ($d=3$) we have
\begin{equation}\label{eq:fqt_compressed_3D}
    f_{\tt mf}(q,t) \propto \exp\left[ -(t/\tau_r)^{3/2} \right]
\end{equation}
which is already giving a possible explanation for the commonly seen exponent 
$\beta \simeq 3/2$ fitted from experimental data when assuming compressed 
exponential decay in time of $f_{\tt mf}(q,t)$.

This idea was taken by Bouchaud \& Pitard to build a mean-field model to predict the 
anomalous relaxation phenomenon~\cite{BouchaudPi-EPJE2001, BouchaudPi-EPJE2002, Bouchaud-InBook2008}.
They have taken into account that the local deformation stops at some point and 
included a new relevant time scale $\theta$, corresponding to the duration of a plastic rearrangement. 
They analytically compute
\begin{equation}\label{eq:fqt_bouchaud}
    f_{\tt mf}(q,t) = \left<\exp\left[i \mathbf{q} \cdot (\mathbf{u}(\mathbf{r},t_0+t)-\mathbf{u}(\mathbf{r},t_0))\right]\right>
\end{equation}
and extract the relevant regimes
\begin{align*}
  f_{\tt mf}(q,t) \propto \begin{cases}
      \exp[-a_b(qt)^{3/2}] & \text{if $t \ll \theta$} \\
      \exp[-a_d(q^{3/2}t)] & \text{if $t \gg \theta$} 
    \end{cases}
\end{align*}
\noindent where $a_b$ and $a_b$ are constants.
The $q^{3/2}$ dependence reflects the fact that the distribution of local displacements 
$u$ decays as $u^{-5/2}$ and therefore has a diverging variance~\cite{Bouchaud-InBook2008}. 
One can see that in the case of very high plastic activity (or temperature), the  
$P(u) \sim u^{-5/2}$ heavy tail is always suppressed, and therefore we recover
the usual $q^2$ dependence in the diffusive regime. 

Notice that 
the mean-field description of Bouchaud \& Pitard~\cite{BouchaudPi-EPJE2001} 
always predicts a purely ballistic-like scaling between space and time $q^{-1}\propto t$ 
in the regime where correlation functions decay as compressed-exponentials. 
Further, the presented mean-field approach predicts no temperature- or $q$-dependency for the compressed exponent. 
In experiments however, temperature- or $q$-dependency are commonly reported and two clear trends appear in the experimental literature:

{\it (i)} On one hand, frequently a $q$-dependence of $\beta$ is 
reported~\cite{DuriCi-EPL2006, FalusPRL2006, RobertEPL2006, CaronnaChMaCu-PRL2008,RutaChMoCiPiBrGiGo-PRL2012,ruta2013} 
and along with that the  $q^{-1}\propto t$ relation is broken.
In particular, the physics is better described by~\cite{CaronnaChMaCu-PRL2008} 
\begin{equation}\label{eq:fqt_compressed_caronna}
    f(q,t) \propto \exp\left[ -c(q^\alpha t)^{\beta} \right].
\end{equation}
Clearly, spatial correlations among plastic events play a role in the relaxation dynamics.
Already, an improvement in theoretical predictions is seen when intermittency of plastic 
events is taken into account ({\it versus} a continuous ballistic process).
In \cite{DuriCi-EPL2006}, Duri {\it et al.} showed that a model of intermittent dynamics
can predict a $q$-dependency of $\beta$.
Using a Poisson distribution for $P_\tau(n)$, the probability that $n$ plastic events 
affect the scattering volume during a time span $\tau$, they are able to
justify their observation of a $\beta$ ranging from from $1.5$ to $1$ in 
the dynamic light scattering of a colloidal gel for increasing $q$. 
The variation of $\beta$ with $q$ was also observed at low temperatures in glass 
forming liquids~\cite{CaronnaChMaCu-PRL2008}. 

{\it (ii)} Another ubiquitous observation is the dependency of the shape exponent
$\beta$ with temperature~\cite{CaronnaChMaCu-PRL2008, RutaPRL2012, LuoPRL2017, AminiPRM2021}.
It has been seen that $\beta$ decreases systematically as temperature is increased.
Furthermore, in~\cite{RutaPRL2012, AminiPRM2021} it's suggested that the 
compressed-exponential behavior of $f(q,t)$ is lost when approaching 
the glass transition temperature $T_g$ estimated from calorimetric measurements of 
the metallic glasses alloys.
Although it's intuitive to assume that the melting of the material leads to a loss of the solid support and therefore elastic 
perturbations induced by localised plastic events cease to exist, it might not be the only reason for loosing the compressed-exponential behavior.
For instance, thermal agitation could start being relevant much before the 
melting of the glassy mixture, and, on the other hand, there are numerical
evidences for the subsistence of Eshelby events in the super-cooled liquid phase~\cite{LemaitrePRL2013, lemaitre2014structural, LemaitreJCP2015}.

In fact, the temperature-dependency of the relaxation remains 
an open issue on the theoretical side.
It has not been discussed in the previously quoted mean-field-like approximations, 
and only recently included in a related analytical approach~\cite{TrachenkoJPCM2021}.
We address such an elasticity-temperature interplay in the present work in 
a fully spatial description.

\section{Model} 
\label{sec:model}

In the last decade, different coarse-grained approaches have been developed to help in 
the understanding of amorphous solids under deformation~\cite{NicolasRMP2018}.
Such lattice-based models known as ``elasto-plastic'' (EP) have the advantage of being capable to 
address larger scale statistics of the dynamical phenomena related with plastic deformation
in amorphous solids, than the one provided by (more ``realistic'') off-lattice particle-based 
models.
If we want to describe a material with a given extension and in a given time-frame, 
the computational demand is, of course, much larger for a particle based approach.
The coarse-graining solves that, allowing us to address length and time scales that involve
several shear-transformation-zones (STZ) to study their interplay and statistics.

Yet an automatic drawback of EP models is the absence of ``particle coordinates'', from which
many common observables are usually defined.
By construction, EP blocks are supposed to describe a {\it region} or patch of the material,
involving several particles, where a STZ or ``plastic event'' can take place. 
A way overcome this limit, while still keeping the enormous functionality of
EP models in the statistical description of amorphous materials is to construct a parallel system of lattice-free
{\it tracer particles}. 
These point-like particles will follow the instantaneous displacement fields derived from
the EP-model, and provide us with virtual particle-coordinates and configurations that we can use to define quantities 
as the mean-square-displacement and the dynamical structure factor.
Assuming that localized relaxation events perturb the surrounding material leading to a stress
redistribution in form of an Eshelby response, one can convolute the simultaneous action
of several plastic events occurring in different locations to workout the displacement field
${\bf u}({\bf r})$ that they induce at any place in the system.
Of course, the resolution of the corresponding displacement field is given by the coarse-graining of the elasto-plastic 
lattice dynamics, but once we derived it we can ignore the the lattice for the movement of tracers.
This construction was used in~\cite{FerreroPRL2014} for the study of a 2D system; we extend 
it here to the case of three dimensions and add a thermal noise to the dynamics of
tracers in a way fully compatible with thermal activation of plastic events in the
background EP model.

\subsection*{A thermally activated and spatially symmetrized elasto-plastic model}

An amorphous material is represented by a coarse-grained scalar stress field 
$\sigma(\br,t)$, at spatial position $\br$ and time $t$ under an externally 
applied shear strain. 
The space is discretiszd in patches or blocks. 
At a given time, each block can either be elastic (``inactive'') or plastic 
(``active'', i.e., locally relaxing). 
This state is defined by the value of a binary variable: $n(\br,t)=0$ for inactive,
$n(\br,t)=1$ for active.
A huge model simplification consist in passing from a fully tensorial description 
to a scalar one, for example, by assigning all the plastic deformation to one scalar 
component of the deviatoric strain. 
While this approach is analytically justified in the case of sheared materials (e.g.,~\cite{JaglaPRE2020}), 
we would need here to take into account the tensoriel caracter since we do not have any symmetry-breaking external applied deformation.
To be able to keep a simple scalar description, we decided to symmetrize the reponse in the three possible shear `directions', not favoring one spatial coordinate over the others.

We define our EP model in $3$-dimensions discretized on a cubic lattice
of $N=L_x\times L_y\times L_z$ blocks, with the stress $\sigma_i$ on a generic 
block $i$ subject to the following evolution in real space
\begin{equation}\label{eq:eqofmotion1}
\frac{\partial \sigma_i(t)}{\partial t} =
  - g_0 n_i \frac{\sigma_i(t)}{\tau} +\sum_{j\neq i} G_{ij} n_j(t)\frac{\tilde{\sigma}_j(t)}{\tau} ;
\end{equation}
where 
$g_0>0$ sets the local stress dissipation rate for an active site, 
the time-dependent local state variable $n_i=\{0,1\}$ indicates if a site is 
undergoing a plastic event (active ($n_i=1$) or inactive ($n_i=0$)),
and the kernel $G_{ij}$ is the Eshelby stress propagator~\cite{PicardAjLeBo-EPJE2004}.
In most EPMs, ${\tilde{\sigma}_j(t)}$ is simply ${\sigma}_j(t)$, 
the stress at site $j$. 
Yet, in this case we have used $\tilde{\sigma}_j(t)=\pm \mu \epsilon_0$, 
where the sign ($\pm$) indicates whether the plastic activation has occurred
in a positive ($+$) or negative ($-$) stress threshold and 
$\epsilon_0$ quantifies a fixed intensity for the Eshelby event
during the plastic event. 
This our choice for ${\tilde{\sigma}_j(t)}$ is justified in 
a subsection below.
Furthermore, notice that typically a $\mu \dot{\gamma}^{\tt ext}$ term
(with $\mu$ the shear modulus and $\dot{\gamma}^{\tt ext}$ the 
externally applied strain rate) appears in the R.H.S., but in our
case it's identically zero (no external driving).

The form of $G$ in $d=3$ can be more easily expressed in Fourier space
\begin{equation}\label{eq:propagator3d_fourier}
G_{\bf q}^{\tt 3D} = -\frac{4q_l^2q_m^2+q_n^2(q_l^2+q_m^2+q_n^2)}{(q_l^2+q_m^2+q_n^2)^2}
\end{equation}
for $\bf q\ne 0$ and, in our scheme of non-conserved stress, 
\begin{equation}
G_{\bf q=0}=-\kappa
\label{app_eshelby_kernel_q0}
\end{equation}
with $\kappa$ a numerical constant set to 1.
The sub-indices $l,m,n$ of the wavevector $\bf q$ components in Eq.~\ref{eq:propagator3d_fourier} are set to different possible 
permutations of the system coordinates $x,y,z$, as explained later.
The last term of (\ref{eq:eqofmotion1}) constitutes a \textit{mechanical noise}
acting on $\sigma_i$ due to the instantaneous integrated plastic activity
over all other blocks ($j\neq i$) in the system.
The elastic (e.g. shear) modulus $\mu=1$ defines the stress 
unit, 
and the mechanical relaxation time $\tau=1$, the time unit of the problem.

\subsubsection*{The thermal activation rule}
The picture is completed by a dynamical law for the local state variable $n_i=\{0,1\}$. 
In the athermally driven case, a \textit{plastic event} occurs in block $i$
($n_i : 0 \rightarrow 1 $), with certain probability~\cite{FerreroSM2019},
when the local stress $\sigma_i$ overcomes a local yield stress $\sigmaY_i$.
In the current work, the driving is absent and is replaced by thermal activation.
When $T>0$ we expect activation to occur with a finite probability
even if $|\sigma_i|<\sigmaY_i$, 
where now we should equally consider the probability of yielding
when a block builds a sufficiently large `negative' stress.
In particular, we use the following rule for sites activation:
$n_i : 0 \rightarrow 1 $ as soon as $|\sigma_i|\geq \sigmaY_i$, or, at any time 
with probability
\begin{equation}\label{eq:activationrulesEPM}
 p_{\tt act}(T) = e^{-B (\sigmaY_i-|\sigma_i|)^\alpha/k_B T}
\end{equation}
\noindent when $|\sigma_i| < \sigmaY_i$~\footnote{Rigorously speaking, 
its a bit more complicated than that, since
we always keep track of both the probability $p^+_{\tt act}(T)$ of yielding 
in the `positive' threshold $\sigmaY_i$ and the probability $p^-_{\tt act}(T)$
of yielding in the `negative' threshold $-\sigmaY_i$.
The complete definition being
$p^\pm_{\tt act}(T) = e^{-B (\pm1(\pm \sigmaY_i-\sigma_i))^\alpha/k_B T}$.}.
We have chosen for simplicity $\sigmaY_i=1$ for all sites (and $k_B=1$). 
We checked that the use of distributed stress thresholds does not change qualitatively our findings.
The factor $B$ can be seen as a measure of facilitation of local 
slips and therefore considered a material-dependent parameter.
In our model, it remains a free parameter that we control.
In the following we set $\alpha=3/2$, taking into account 
the discussion in~\cite{FerreroSM2019, FerreroPRM2021} which 
identifies it as the exponent expected for smooth energetic barriers
and the most comparable one with atomistic simulations and 
experiments~\footnote{We have checked nevertheless that the use of
other values of $\alpha$, $\alpha=1,2$, does not affect our 
conclusions.}.
Finally, an active block $i$ becomes inactive $n_i :  0 \leftarrow 1$
with a constant probability $\tau_{\tt ev}^{-1}$.
The prescribed time $\tau_{\tt ev}$ is the  \textit{`lifetime'} of an active event 
and the previous stochastic rule guarantees that, on average, plastic 
events have such a duration.
The value for $\tau_{\tt ev}$ is also not fixed and remains a free model parameter.

\subsubsection*{Symmetrized elastic propagator and strain events}
In the absence of an external shear, there's no preferred \textit{direction} 
for an Eshelby event.
In principle we should consider that the local shear transformations occur 
arbitrary orientations and angles.
This apparently ingenuous statement can largely complicate the numerical
implementation of the model and it's unnecessary for our purposes.
Even though a bit less realistic, for simplicity we have chosen to preserve the
symmetry only between the three principal shear planes of our $d=3$ sample 
geometry ${xy},{yz},{zx}$.
We define not one but three different propagators related to these shear planes: 
Eq.~\ref{eq:propagator3d_fourier} with three index permutations for $(l,m,n)$:
$(x,y,z)$, $(y,z,x)$ and $(z,x,y)$.
When the criterion for local yielding is met, one shear plane is chosen
randomly and that site will be shear-transforming in that orientation only during
it's activity period.
The next activation of the same site can occur in a different orientation.
%
In this way we maintain a single local scalar variable representing the stress on each block, 
despite the introduction of various possible orientations of a plastic event. 
Moreover, the absence of an externally applied shear restrains the system
to plastic activity only induced trough a \textit{`thermal bath'}.
Local stresses are close to Gaussian-distributed around zero 
and the width of the distribution increases with $T$ (see App.~\ref{app:stress_distributions}).
In a sense, our system enters after a transient dynamics depending on the initial state, always in thermal equilibrium.
In the absence of externally applied deformation, we associate to a plastic event a characteristic strain rather than a 
characteristic stress~\cite{FerreroPRL2014} as is usually the case for EPMs with driven dynamics. 
As mentioned, in practice $\tilde{\sigma}_j(t)$ in Eq.(\ref{eq:eqofmotion1}) 
is defined as 
\begin{equation}\label{eq:sigma_tilde}
  \tilde{\sigma}_j(t)=\pm \mu \epsilon_0,  
\end{equation}
if the site is active,
where the sign ($\pm$) depends on the plastic activation occurring
at a positive ($+$) or negative ($-$) stress threshold.
$\epsilon_0$ constitutes a parameter of our model and quantifies the 
intensity of the Eshelby event; it can be seen as the $\epsilon V_1$ 
in Eq.(\ref{eq:displdipole}).

\subsubsection*{Displacement fields and tracer particles}

The yielding of a block in different shear planes will give rise to displacement
fields in the rest of the system, which we capture using the Oseen tensor
components, following and generalizing~\cite{PicardAjLeBo-EPJE2004}.
In Fourier space

\begin{equation}\label{eq:displ_field}
    \hat{{\bf u}}({\bf q}) = \hat{\mathbf{O}}({\bf q}) \cdot (2i \mu {\bf q} \cdot 
    \hat{{\bf \epsilon}}_{\tt pl})
\end{equation}

\noindent ${\bf u}({\bf r})$ (the real-space counterpart of ${\bf u}{(\bf q})$) 
is the vectorial displacement field, 
${\bf \epsilon_{\tt pl}}({\bf r})$ is a plastic strain shear-component
($\epsilon_{\tt pl}{}_j = n_j \tilde{\sigma}_j/\mu$, the strain related to the stress multiplying the propagator 
in Eq.~\eqref{eq:eqofmotion1}), and $\mathbf{O}({\bf r-r'})$ is the 
translationally-invariant Oseen tensor.

\begin{equation}
    \hat{\mathbf{O}}({\bf q}) = \frac{1}{\mu q^2} \left(\mathbf{I} - \frac{{\bf q}{\bf q}}{q^2} \right)
\end{equation}

Given an instantaneous configuration of the system's plastic activity 
(with its complexity of three possible local yielding directions for each block),
one can convolute the different contributions given the active blocks 
to the displacement field at \textit{any} position in the system by using
Eq.(\ref{eq:displ_field}) in Fourier and then transforming ${\bf u}{(\bf q})$ 
to real space.
Here we consider for the shear components of $\hat{\mathbf{O}}({\bf q})$ 
the same three possible permutations that we used for the propagator $G_{\bf q}$.
Notice that for a more detailed description one would need to be careful about the 
displacements in an event’s core region, which follow a different (exponential) decay~\cite{WangPRE2022}.
Here we had simply taken the precaution of setting things such that a 
plastic event has a null effect in the displacements of tracers that 
are transiting its own cell and only affect the tracers outside the 
event's core.

With this, tracer particles are simulated in parallel to our EP model 
evolution, simply following the displacements fields.
A set of $M$ probe particles initially located at random in the
cube $(L_x,L_y,L_z)$ simply follow the EP-model-generated displacement fields
updating the position ${\bm \xi}_s$ for tracer $s$ by

\begin{equation}\label{eq:motion-tracers}
    {\bm \xi}_s \to  {\bm \xi}_s + \mathbf{u({\bm \xi}_s)} dt
\end{equation}
\noindent on its three scalar components.
We call these \textit{`athermal tracers'}, since temperature does not step 
in explicitly in their dynamics.
This is modified in Sec.~\ref{sec:fully_thermal} when we include a thermal 
noise acting on the tracers.

From the movement of these tracer particles we will compute most of our
quantities of interest. 
In a real system, the particles that we observe are in general not tracers
but the system constituents themselves.
Therefore, for example, dilation occurring inside the STZs will contribute to
particle displacements~\cite{LuPRA2018}.
Our approach disregards those contributions and takes only into account displacements generated by the long-range elastic response. 

\section{Results}
\label{sec:results}

\begin{figure}[t!]
\includegraphics[width=0.95\columnwidth]{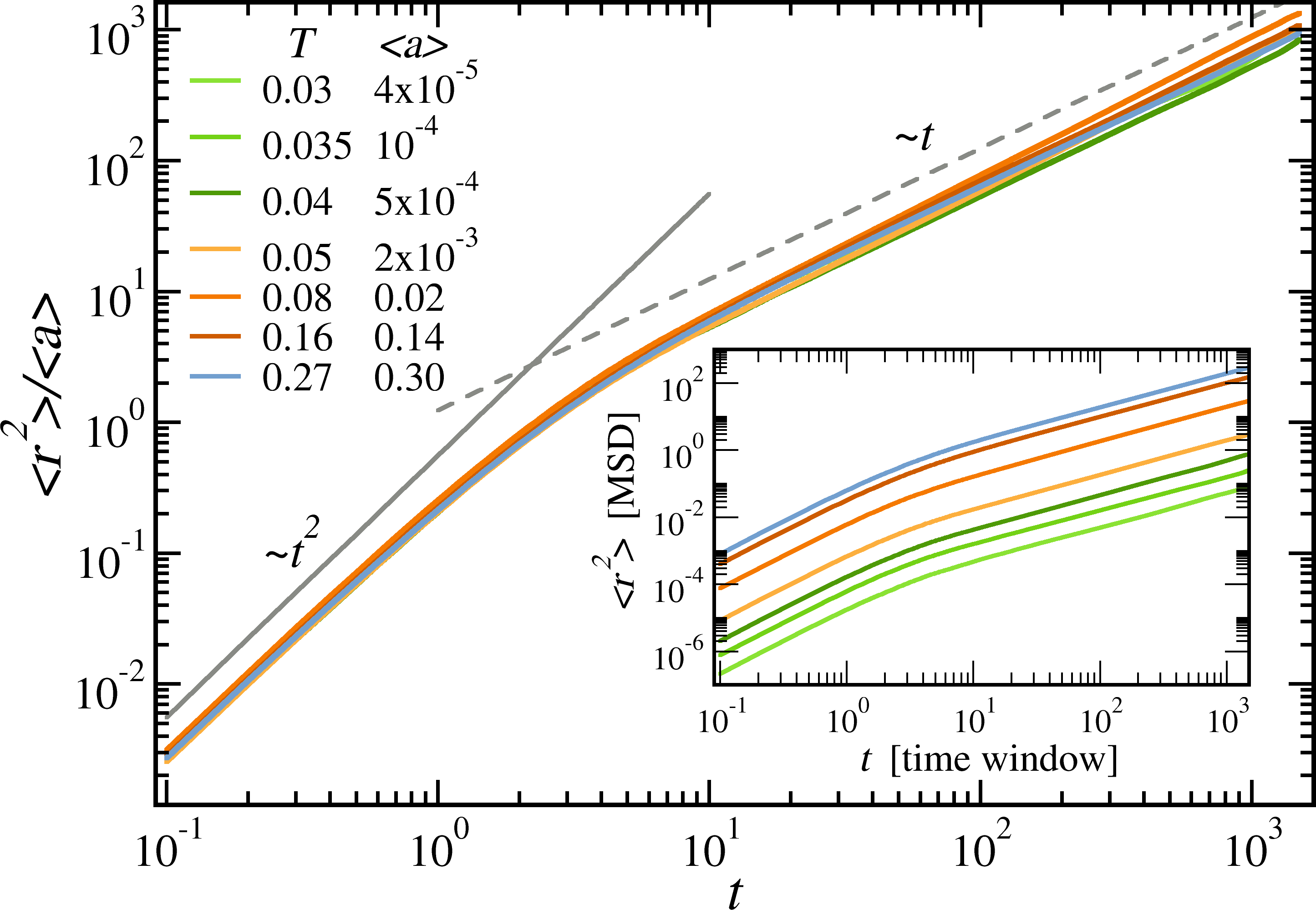}
\caption{\label{fig:msd_athermal_tracers} 
{\it Mean square displacement as a function of observation time window}.
The main plot shows the MSD $\left<r^2\right>$ rescaled by the mean 
activity $\left<a\right>$ for seven different activities/temperatures
as indicated in the labels,
The gray solid and dashed lines are guidelines to show the ballistic ($\sim t^2$) and 
diffusive ($\sim t$) behaviors.
The inset shows $\left<r^2\right>$ unscaled.
The plastic events duration is $\tau_{\tt ev} = 1.5$. $\epsilon_{0} = 1.0$. $L = 32$.}
\end{figure}

In this section we present the numerical results for the relaxational 
dynamics of our quiescent EP model at finite temperature.
All results correspond to {\it steady-states}, which due to the absence 
of external driving can be considered as being {\it equilibrium states}, 
as discussed in~\cite{FerreroPRL2014}.
In a real material at rest (quiescent), the plastic activity might decrease 
(and even stop at some point) when the residual stresses of the material 
preparation are exhausted. 
Here, we perform measurements at fixed temperatures or fixed plastic activity 
levels and do not consider any material aging in the long term.
We therefore expect our results to be comparable to the cases in which a given 
level of plastic activity is maintained in experiments or atomistic simulations
during the measurements.

\subsection{Thermally induced plastic events for `athermal' tracers}
\label{sec:athermal}

We start by discussing the extension to three dimensions of the results presented
in~\cite{FerreroPRL2014}.
A finite temperature rules the rate of plastic activity resulting in displacement of
tracer particles, but those tracers move just according to the displacement fields
without any other perturbing force.

\subsubsection{Mean square displacement}

Figure~\ref{fig:msd_athermal_tracers} shows for different temperatures
the mean square displacement (MSD) defined as 
$\left<r^2\right>~\equiv~\overline{\frac{1}{N}\sum_i r^2_i}$, where 
$r_i~=~|{\bf r}_i(t_0+ t)-{\bf r}_i(t_0)|$ is the distance traveled by 
tracer $i$ in the time-lapse $t$ and the over-line indicates an extra 
average using sliding time windows, by moving $t_0$ in the stationary state.
Tracer particles move following the vector displacement 
field ${\bf u}(r,t)$ (Eq.~\ref{eq:displ_field}).
Data corresponds to cases of seven different mean plastic activities
$\left<a\right> \simeq [4\!\times10^{-5}, 1\!\times10^{-4}, 5\!\times10^{-4}, 2\!\times10^{-3}, 0.02, 0.14, 0.3]$ resulting from temperatures 
$T \simeq [ 0.03,\, 0.035,\, 0.04, \,0.05, \,0.08, \,0.16, \,0.27]$, respectively, and $B=1$.
For a given temperature, we observe that $\left<r^2\right>$ behaves {\it ballistic-like} 
($\sim t^2$) if the time window observation is small, and, {\it diffusively} 
($\sim t$) for larger time windows.
The characteristic time scale separating this two regimes is the duration of the
plastic events $\tau_{\tt ev}$ 
(which for data in Fig.~\ref{fig:msd_athermal_tracers} is $\tau_{\tt ev} = 1.5$). 
What is identified as ``ballistic'' here is nothing but a regime dominated by
the persistent motion of tracer particles during the elastic response in a given direction each, with  
little or none deviation.
Beyond the persistent time controlled by $\tau_{\tt ev}$, particles diffuse.
As expected, larger plastic activities lead to larger plastic displacements and 
larger effective diffusion coefficients $D_{\tt eff} \equiv \frac{\left<r^2\right>}{6t}$
at long $t$.
In the main-plot of Fig.~\ref{fig:msd_athermal_tracers} we are able to collapse all curves 
when normalizing the MSD by the mean plastic activity $\left<r^2\right>/\left<a\right>$, 
where $\left<a\right>= \left<\frac{1}{N} \sum_i n_i\right>$ and $\left< \cdot \right>$
denotes average over time. 

It's worth clarifying something again about the persistent regime observed here.
In the relaxation of glasses, a so-called $\beta$ relaxation takes place
first, inside the cages formed by the frustrated material. 
At times smaller than the beta relaxation time the mean-square displacement can 
display a true ballistic regime where particles move freely within their cage.
EPMs do not catch such a microscopic dynamics, they are defined at a 
mesoscopic level.
All the relaxation presented in this work should be related to a relaxation 
occurring far beyond the $\beta$ relaxation, at much larger time and length scales. 
The tracer displacements we consider are the result of the elastic response
to plastic events that in real systems can involve up to a few dozens of particles. 
As already suggested in~\cite{FerreroPRL2014}, we believe that the EPM approach catches 
the relevant length scales where compressed relaxation is observed in experiments. 
And by that we mean the regime of wave vectors comparable to the typical core size of 
a local relaxation event in the material 
and at times related to the typical duration of those events.

\subsubsection{Dynamical structure factor}

\begin{figure}[t!]
\includegraphics[width=0.9\columnwidth]{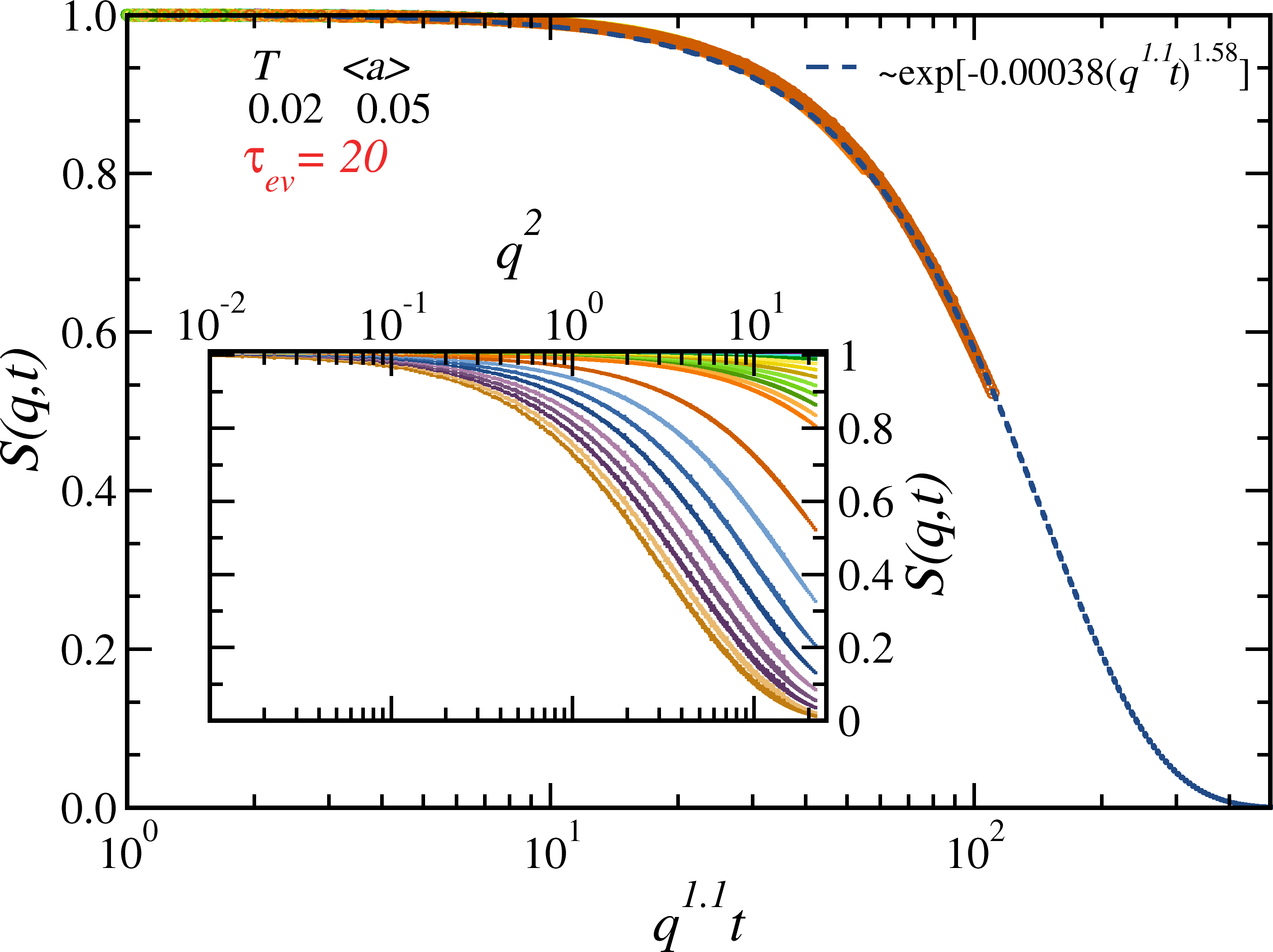}
\caption{\label{fig:Sq_athermal_ballistic} 
{\it Dynamical structure factor $S(q,t)$ relaxation for low plastic activity
and short times (in the ballistic regime)}.
The inset shows $S(q,t)$ as a function of $q^2$ for time windows up to $t=100$,
while the duration of plastic events is $\tau_{\tt ev} = 20$. 
The mainplot collapses the curves corresponding to time-windows $t \leq \tau_{\tt ev}$,
plotting $S(q,t)$ as function of $q^{1.1}t$.
A compressed exponential behavior $S(q,t) \propto \exp \left[-A( q^{1.1}t)^\beta\right]$  is fitted with $\beta\simeq 1.58$ (blue curve).
$T=0.02$. 
$\epsilon_0 = 0.1$. $L = 32$.
}
\end{figure}

We further compute the main observable of the relaxation dynamics,
the dynamical structure factor $S(q,t)$, analogous of Eq.~\ref{eq:fqt_bouchaud}:
\begin{equation}\label{eq:sqt}
    S(q,t)= \frac{1}{M}\left \langle \sum_{n=1}^{M}\cos\left[\mathbf{q}.\left(\mathbf{r}_{n}(t+t_{0})-\mathbf{r}_{n}(t_{0})\right)\right] \right \rangle,
\end{equation}
which is a measure of the decorrelation of tracer particles positions in time
respect to an initial configuration.
Here $M$ is the total number of tracer particles, and the brackets indicate a sliding time-window average with different $t_{0}$ and the different discretized wave vectors
$\mathbf{q}$ that share the same modulus $q$.
In fact, we find it easier to get averaged curves at fixed times $t$ rather than 
at fixed $q$ values.
It's worth mentioning that although we are computing a self-ISF in the
definition of Eq.\eqref{eq:sqt}, from tracers that do not interact with each other,
they all follow displacement fields that are a result of the collective behavior 
of plastic events~\footnote{We expect our description to be compatible with one where 
the XPCS is observed at ``large'' values of $q$, on a length-scale 
compatible with the size of the STZ.
In fact, it was suggested that ISF measured at the peak of the structure factor 
reflects the behavior of the far fields of the van Hove correlation function~\cite{WuPRL2018}. 
In the EPMs this comes true by construction with the Eshelby propagator.}.

Figure~\ref{fig:Sq_athermal_ballistic} shows the dynamical structure factor for curves
corresponding to short times ($t \lesssim \tau_{\tt ev}$). 
This means that the data displayed is collected from the persistent tracer movement regime. 
To observe such regime clearly we have set a large event duration, 
$\tau_{\tt ev}=20$, which in turn enforces a reduction of the 
parameter $\epsilon_0$ in Eq.\ref{eq:sigma_tilde} in order to maintain 
the mean activity at low levels ($\left<a\right> = 0.05$).
Here $\epsilon_0 = 0.1$ has been used. %
A collapse of different curves can be seen when we plot $S(q,t)$ vs $(q^{1.1}t)$,
and  $S(q,t)$ presents the shape $S(q,t)\!\propto\!\exp\left[-A\left(q^{\alpha}t\right)^\beta\right]$, 
with $\alpha \simeq 1.1$, $\beta\simeq 1.58$.
In this short time regime therefore we observe: 
(i) $\tau_r \sim q^{-1}$, typical of ballistic processes, and 
(ii) a compressed shape exponent $\beta$ in the range expected from 
experiments~\cite{cipelletti-2000} and close but different from 
mean-field theory~\cite{BouchaudPi-EPJE2001,Bouchaud-InBook2008}. 

\begin{figure}[t!]
\includegraphics[width=\columnwidth]{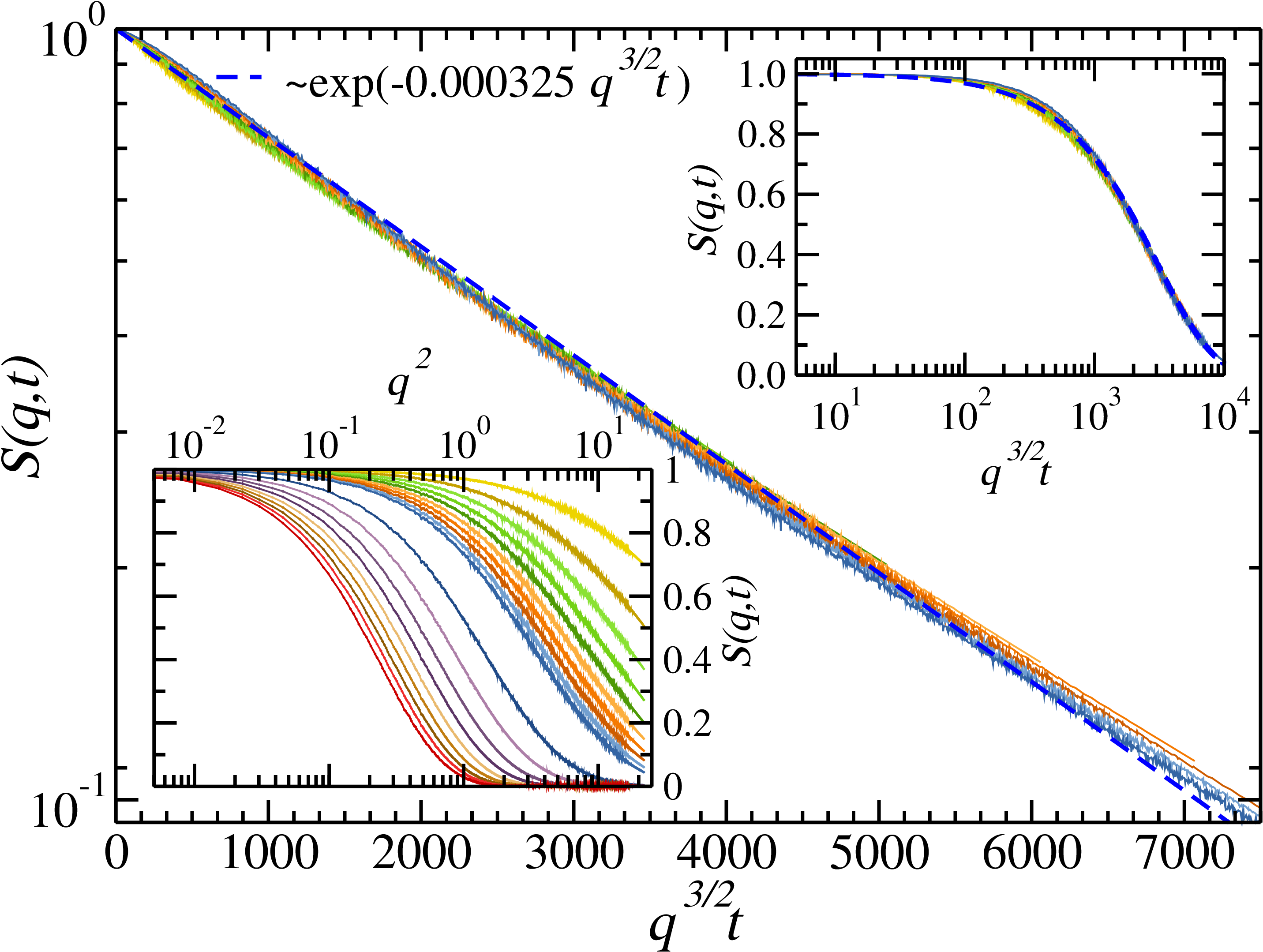}
\caption{\label{fig:Sq_athermal_diffusive} 
{\it Dynamical structure factor $S(q,t)$ relaxation for low plastic activity
and long times (in the $q^{3/2} t$ diffusive regime)}.
The lower-inset shows $S(q,t)$ as a function of $q^2$ for time windows up to $t=10000$.
The duration of plastic events is $\tau_{\tt ev} = 1.5$. 
The mainplot and upper-inset collapse the curves corresponding to time-windows 
$ \tau_{\tt ev}\leq t\leq 1000$, plotting $S(q,t)$ as function of $q^{3/2}t$.
A pure exponential behavior $S(q,t) \propto \exp \left[-A(q^{3/2}t)\right]$ is 
fitted.
$T=0.04$. $\left<a\right> \simeq 0.0005$. $\epsilon_{0} = 1.0$. $L= 32$.
}
\end{figure}

\begin{figure}[b!]
\includegraphics[width=\columnwidth]{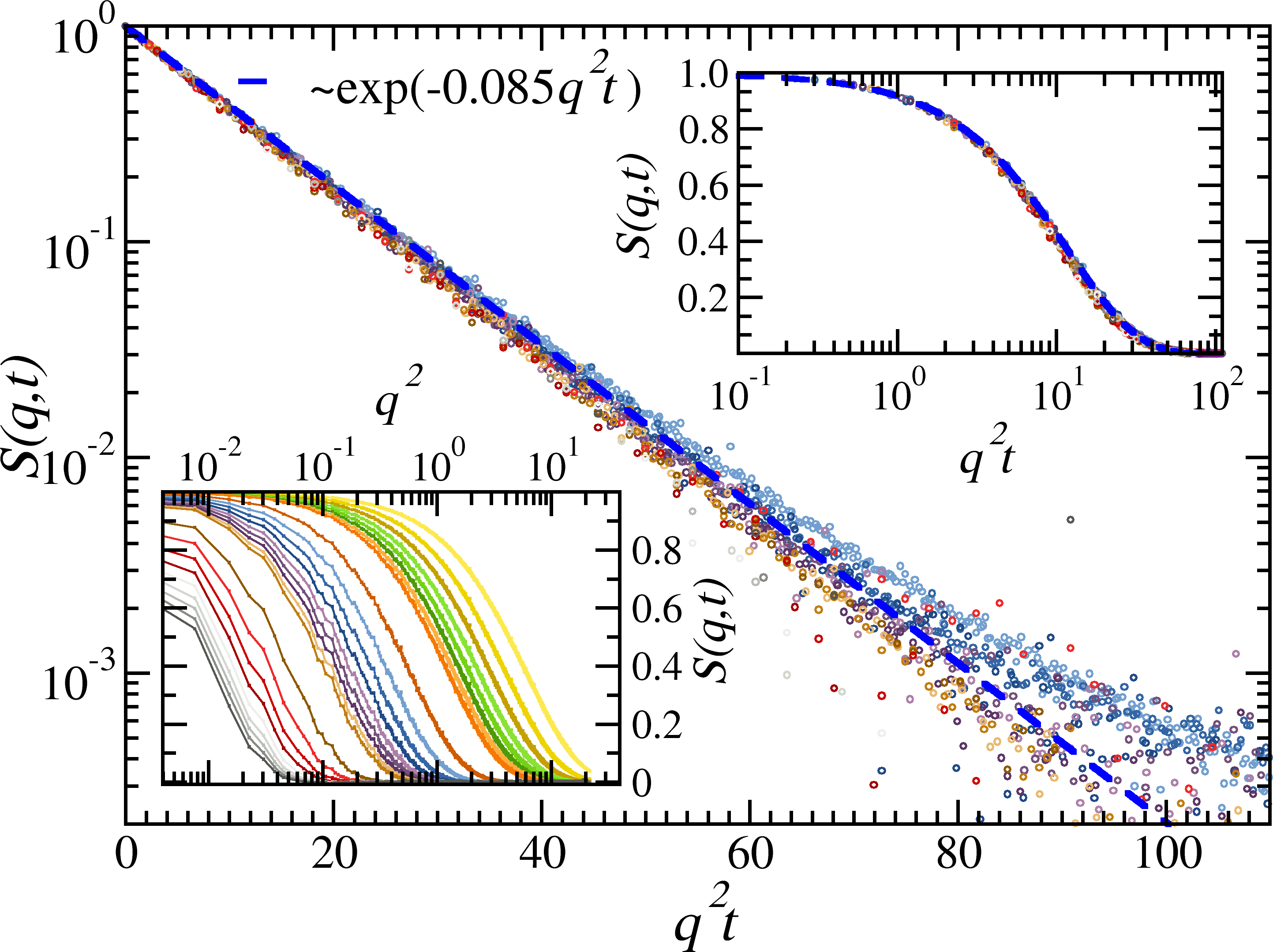}
\caption{\label{fig:Sq_purediffusive} 
{\it Dynamical structure factor $S(q,t)$ relaxation for moderate/high plastic activity 
and long times (in the usual $q^2 t$ diffusive regime)}.
The lower-inset shows $S(q,t)$ as a function of $q^2$ for time windows up to $t=1000$.
The duration of plastic events is $\tau_{\tt ev} = 1.5$. 
The mainplot and upper-inset rescale the curves corresponding to time-windows $t>20$ 
(curves on the left of the dark-orange one in the inset), plotting $S(q,t)$ as function of $q^{2}t$.
A pure exponential behavior $S(q,t) \propto \exp \left[-A(q^{2}t)\right]$ is 
fitted.
 $T=0.16$. $\left<a\right> \simeq 0.14$. $\epsilon_{0} = 1.0$. $L= 32$. 
}
\end{figure}

In Figure.~\ref{fig:Sq_athermal_diffusive} we show the dynamical 
structure factor for curves corresponding to long times ($t \gg \tau_{\tt ev}$). 
Without lack of generality, here we have used the data for $\tau_{\tt ev}=1.5$ 
to show comparatively much larger times.
At observation windows $t \gg \tau_{\tt ev}$, we always expect to observe diffusion.
Here, we see a collapse in the curves for times up to $t=1000$ when we 
plot $S(q,t)$ as a function of $q^{3/2}t$.
The shape of the relaxation is now a simple exponential ($\beta = 1$)
$S(q,t)~\propto~\exp\left[-A\left(q^{3/2}t\right)\right]$.
This is consistent with the mean-field prediction~\cite{BouchaudPi-EPJE2001}:
For $d=3$, $\tau_r \sim q^{-3/2}$ is expected for a diffusive process~\cite{BouchaudPi-EPJE2001,Bouchaud-InBook2008} 
(see Eq.\ref{eq:fqt_compressed_3D} for the $\tau_r$ definition). 
Yet, at much longer times the $\tau_r \sim q^{-3/2}$ scaling breaks down.
As explained by Bouchaud~\cite{Bouchaud-InBook2008}, 
$q^{3/2}$ reflects the fact that the `distribution of local displacements' 
$P(u)$ has a diverging variance. 
For a distribution with a finite variance, one would recover the usual $q^2$ 
dependence in the exponential relaxation.
As we show in the next section, $P(u)$ always has an upper cutoff. 
Therefore, it's not surprising that dynamical structure factors measured over 
very long time windows already sense that finite variance and deviate from the 
$\tau_r \sim q^{-3/2}$ scaling.
Moreover, the $u^{-5/2}$ tail is completely suppressed at higher plastic activities.
This is a simple consequence of the central limit theorem. 
The response in the high activity regime is composed of a sum of a large number of random variables drawn from a distribution, 
which presents a cutoff at large displacements due to the finite core size of the plastic events.
In this case, the $S(q,t) \sim \exp[-Aq^{3/2}t]$ regime completely shrinks
to give place to a ``purely diffusive'' regime $S(q,t) \sim \exp[-Aq^{2}t]$,
as the one observed in Brownian motion, already at time-windows 10 times larger 
than $\tau_{\tt ev}$.
This is shown in Fig.~\ref{fig:Sq_purediffusive}, where we use
$\tau_{\tt ev} = 1.5$ and $T=0.16$ ($\left<a\simeq 0.15\right>$).

\subsubsection{Displacements distribution}
\label{sec:Pofu}

As already mentioned, another quantity of interest
is the characteristic displacement $u$ of a tracer particle in a given time window.
More generally, we are interested in its distribution $P(u)$ at different temperatures.

In the limit of low temperatures (small plastic activities), we expect these
displacements to be ruled by the fields generated by a few plastic events.
We know that a plastic event induces a displacement $u \sim 1/r^{d-1}$, like a dipole.
Close to a plastic event tracers will move a lot, large $u$, but statistically there
will be more tracers further away from the event.
More quantitatively, the probability of `seeing' a plastic event at a distance
between $\left[r, r+dr\right]$ when sitting on a random tracer will be proportional 
to $p(r) \sim r^{d-1} dr$ in general dimension $d$.
Then, using probability conservation and $u \sim r^{-(d-1)}$ ($du \sim r^{-d} dr$)
\begin{eqnarray}
    P(u)du & = & P(r)dr \\ \nonumber
    & \propto & r^{d-1} dr \\ \nonumber
    & \propto & u^{-(2d-1)/(d-1)} du
\end{eqnarray}
Therefore, for large displacements we expect $P(u)$ to decay as $P(u) \sim u^{-(2d-1)/(d-1)}$, $P(u) \sim u^{-5/2}$ in $d=3$.

On the other hand, the statistics of small displacements will be ruled by an
incoherent superposition of small `kicks' given by distant plastic events.
In a given finite system (and with periodic boundary conditions) one cannot be 
further than a maximum distance controlled by the density of events, 
from the closest plastic event.
Then, it's more frequent to find intermediate displacement values; since there are 
many more tracers at intermediate distances from the plastic events than far away.
Indeed, one can prove that $P(u)$ should go to zero as $u$ decreases 
as $P(u) \sim u^{d-1}$.
From just a pure incoherent Brownian motion and the resulting Maxwellian distribution 
for the displacements in a given time window, we are able to derive such $\sim u^{d-1}$
behavior at small $u$ for general $d$ (see Appendix. \ref{app:Pu}).

Finally, in the limit of very low temperatures, the small displacements are not ruled 
by the incoherent superposition of many small kicks, but instead by the 
displacements field generated a very distant single (or few) plastic event(s).
In this case we can expect $P(u)\!\sim\!u$ at small $u$.

\begin{figure}[t!]
\includegraphics[width=\columnwidth]{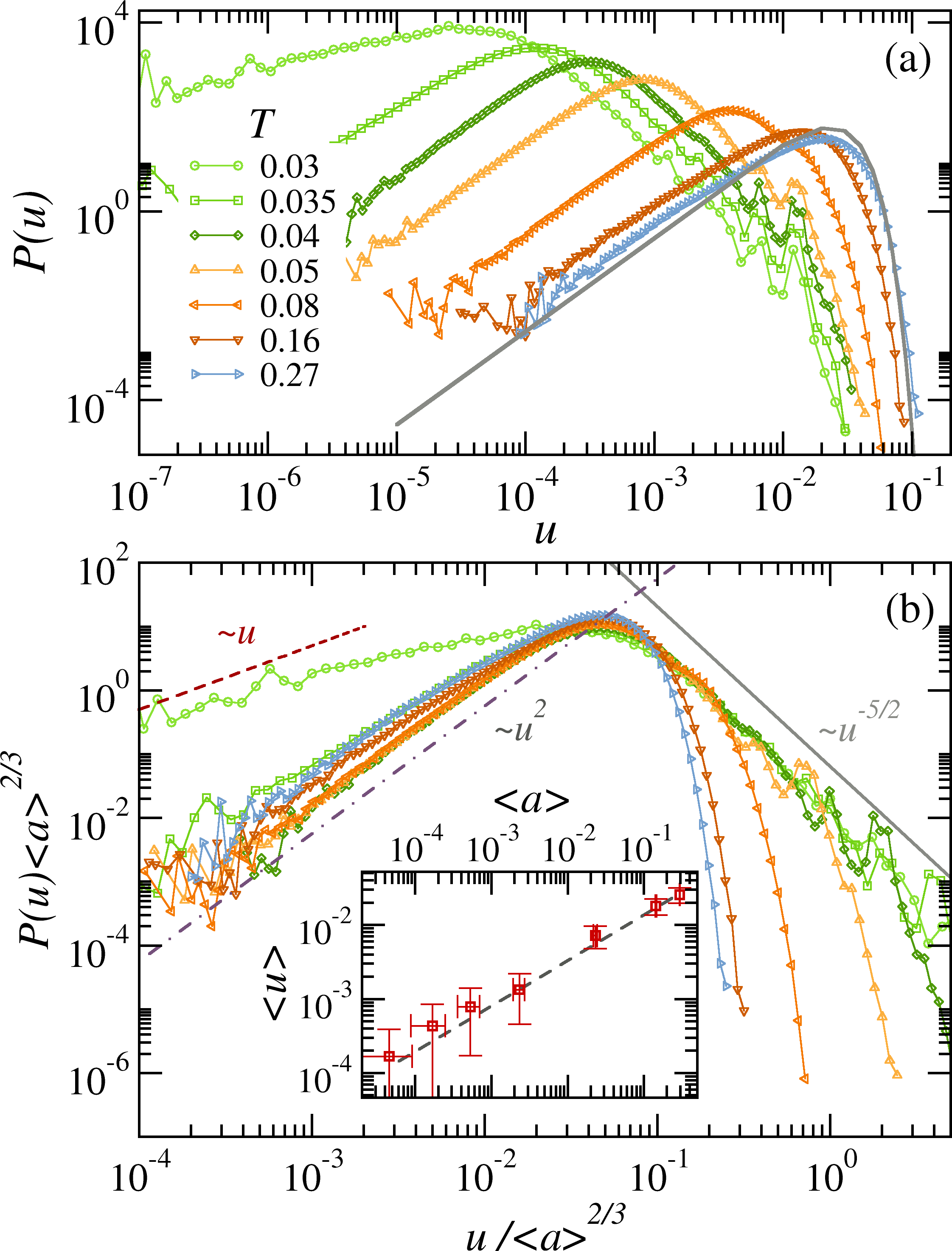}
\caption{\label{fig:Pofu_athermal} 
{\it Displacement distribution $P(u)$ for different temperatures}.
The tracers displacements absolute values $u$ are measured each $0.1$ time units.
Panel (a) shows $P(u)$ for different temperatures 
$T= 0.03, 0.035, 0.04, 0.05, 0.08, 0.16$ and $0.27$ 
in different colors and symbols.
The solid gray line correspond to the analytical expression of Eq.~\ref{eq:maxwell_distribution} for $c=0.016$.
The inset of panel (b) shows the mean displacement $\left<u\right>$ 
vs mean activity $\left<a\right>$. 
The dashed line is a power-law $\sim \left<a\right>^{2/3}$.
Panel (b) shows rescaled curves $P(u)\left<a\right>^{2/3}$ vs. 
$u/\left<a\right>^{2/3}$. 
Red dashed-line, purple dot-dashed-line and gray solid line 
show $P(u)\sim u$, $P(u)\sim u^{2}$ and $P(u)\sim u^{-5/2}$, respectively.
$\tau_{\tt ev} = 1.5$. $\epsilon_{0} = 1.0$. $L = 32$.
}
\end{figure}

In Fig.~\ref{fig:Pofu_athermal}(a) we show the displacement distribution $P(u)$,
with $u$ defined as the absolute displacement in a time window $\Delta t = 0.1$.
These distributions have been obtained from $N_d \approx 1.1\times 10^{9}$ 
independent displacements of $M$ particles in the steady state.
Let us first notice that the maximum of $P(u)$ moves to the right as the 
temperature increases. 
This is consistent with a larger average displacement value for tracers at larger
and larger mean plastic activities. 
Which is something to be expected, since kicks coming from several neighboring events
can add up to produce a larger displacement. 
We observe in particular that $\left< u \right> \propto \left< a \right>^{2/3}$
(Fig.~\ref{fig:Pofu_athermal}(b) inset).
The power-law cutoff at large $u$ is eventually controlled by the way
the displacement field is computed on a lattice and the time integration of the displacement set up a natural maximum value. 
At each time step the displacement field cannot be higher than the one felt on a nearest-neighbor cell to an active plastic event. 
That kick times the $\Delta t$ of the displacement observation makes the 
maximum $u$, here $\sim 0.1$.
Appendix~\ref{app:Pu} explores the dependence of $P(u)$ on $\Delta t$.

In Fig.~\ref{fig:Pofu_athermal}(b) we use the 
$\left< u \right> \propto \left< a \right>^{2/3}$ scaling to `align' curves
corresponding to different temperatures.
In this scaled plot, we can appreciate clearly the power-law decay of $P(u)$ 
for large displacements $\sim u^{-5/2}$; which extend to larger and larger 
ranges as temperature is lowered.
The power-law behavior is disrupted at larger activities by the incoherent
superposition of different plastic events effect on displacement fields. 
In fact, one can prove that for very high activities, where plastic events 
are basically uncorrelated, and induce an effective Brownian motion in 
tracer particles the distribution $P(u)$ takes the form of a Maxwellian 
(see gray line in Fig.~\ref{fig:Pofu_athermal}(a)).
The systematic little `peaks' or oscillations observed in the distributions 
tails can be attributed to a discretization effect: 
integrating the tracer displacement over a field defined at patches makes
some values of $u$ {\it a priori} more probable to obtain that others
The choice of a larger time window to define $u$ washes out those oscillations
(see App.~\ref{app:Pu}), so would do a smoothing of the 
displacement field (e.g., by interpolation).

\subsection{Fully thermal system}
\label{sec:fully_thermal}

Up to now we have only considered tracer particles that follow displacement 
fields but that are themselves insensitive to temperature.
We now turn to the slightly more realistic case in which we include thermal 
agitation in the tracer particles equations of motion.
For each tracer $s$ we now update the tracer positions according to

\begin{equation}\label{eq:motion-tracers_therm}
    {\bm \xi}_s \to  {\bm \xi}_s + \mathbf{u({\bm \xi}_s)} dt + \eta_{s}(t,T)\sqrt{dt}
\end{equation}

\noindent with $\mathbf{u({\bm \xi})}$ being the instantaneous displacement field at site ${\bm \xi}$, computed from 
the plastic activity (\ref{eq:displ_field}) and $\eta_{i}(t,T)$ a Langevin thermal noise with zero 
mean value $\left<\eta_{i}(t,T)\right>_t=0$ and delta correlated $\left<\eta_{i}(t,T)\eta_{i'}(t',T)\right>=2T\delta(i-i')\delta(t-t')$. 
The motivation to include the last term in the tracers' position updates comes 
from the fact that in thermal amorphous solids (metallic glasses and 
some colloidal glasses and gels~\cite{NicolasRMP2018,FerreroPRM2021}) 
Brownian motion is relevant.
It does not pretend to replace the complete atomistic modeling of 
interacting particles in a thermal bath, but add enough ingredients
to sense the interplay of plasticity and temperature in a relaxing
glass.
Also, it turns into a closer model representation of the nanoparticles
system presented in~\cite{CaronnaChMaCu-PRL2008}.

Note that this approach will only stand valid for small enough temperatures, 
such that the typical thermal displacements remain much smaller than the linear 
size of the thermally activated rearranging regions (cells of the EPM).
Assuming such low temperatures is anyhow more opportune and realistic, 
since the compressed exponential relaxation is only expected in this 
regime~\cite{Gado2017}.
With respect to Sec.~\ref{sec:athermal} where only the ratio $B/(k_BT)$ was 
relevant (and $B=1$ was used), now the absolute value of $T$ also matters.
In the following, we work with smaller values of $T$, but staying in a 
comparable range of $B/(k_BT)$ by decreasing $B$ accordingly.

In particular, we start by varying the temperature in the tracer's e.o.m. while
fixing the elasto-plastic model mean activity.
This ``fixed $\left<a\right>$ level'' protocol would better mimic the cases 
in which plastic activity is considered to be largely controlled
by rearrangements originated in pre-stress~\cite{Gado2017, SongPNAS2022}.
To do that, we fix the ratio $B/(k_BT)$ in Eq. \ref{eq:activationrulesEPM}.
For most of the Section we will set $\left<a\right> \simeq 0.05$,
which, for comparison, correspond to a temperature $T\simeq 0.1$ when $B=1$, 
$\epsilon_{0}= 1.0$ and $\tau_{\tt ev}$ = 1.5.

\subsubsection{Mean square displacement at fixed activity}

\begin{figure}[t!]
\includegraphics[width=\columnwidth]{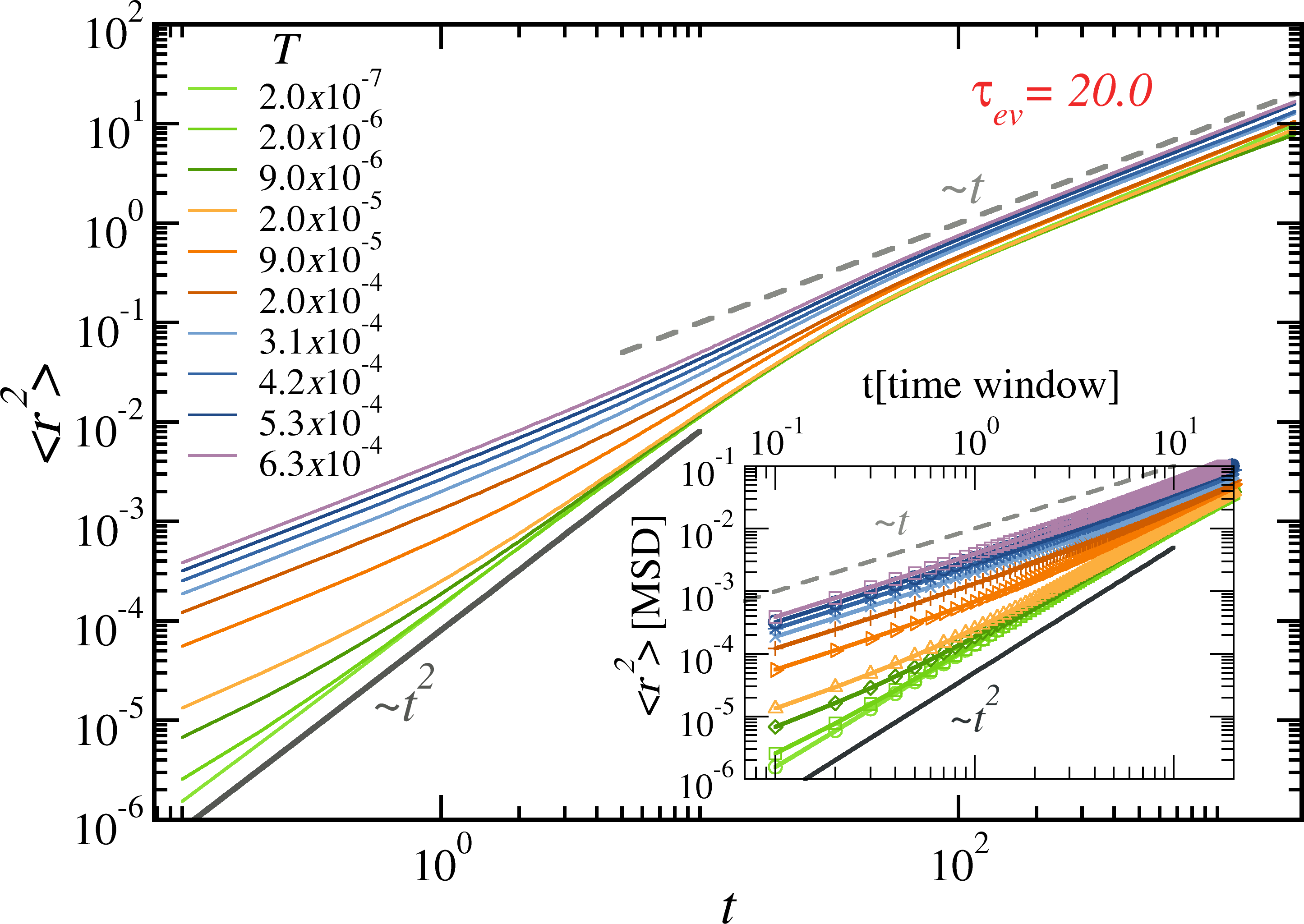}
\caption{\label{fig:msd-thermal-fixed-ratio} 
{\it Mean square displacement $\left< r^{2} \right>$ for a fixed plastic activity.}
The main plot shows the mean square displacement as a function of time observation window for different temperatures, ranging from $T = 2.0 \times 10^{-7}$ 
(light-green curve) to $T = 6.3 \times 10^{-4}$ (light-purple curve).
The prefactor $B$ in Eq.~\ref{eq:activationrulesEPM} has been varied for each temperature such that $\left<a\right>(B/k_{B}T) \simeq 0.05$ is kept fixed.
The gray full- and dashed-lines are guidelines to show the ballistic 
($\sim t^{2}$) and diffusive ($\sim t$) behaviors, respectively. 
The inset shows a zoom-in to the short times regime ($t \leq \tau_{\tt ev}$). 
$\tau_{\tt ev} = 20.0$. $\epsilon_{0} = 0.1$. $L=32$, .
}
\end{figure}

Figure~\ref{fig:msd-thermal-fixed-ratio} shows the mean square displacement 
of tracers at different temperatures and a fixed plastic activity
at a low value $\left<a\right> \simeq 0.05$. 
At low temperatures, we observe the same crossover from a ballistic or persistent
movement regime ($\left<r^2\right> \sim t^2$) to a diffusive regime ($\left<r^2\right> \sim t$) at $t \simeq \tau_{\tt ev}$ as reported in Fig.\ref{fig:msd_athermal_tracers}.
Notice that now we show the case of $\tau_{\tt ev}$ for a better display of the 
`short times' ($t < \tau_{\tt ev}$) regime. 
As we increase the temperature, and thermal agitation becomes more and more
relevant, we can observe how the ballistic regime is washed-out. 
While this is intuitive and somehow unsurprising, it hasn't been addressed 
before (Ref.~\cite{FerreroPRL2014} worked only with athermal tracers and
the mean-field approach of Pitard\&Bouchaud didn't include a thermal noise
either).
Explicitly modeling the interplay of plastic activity and temperature on 
tracer's motion allows us to quantify the emergent dynamics.
For example, one can notice that at intermediate temperatures in 
Fig.~\ref{fig:msd-thermal-fixed-ratio} the MSD shows an "s" shape. 
Indeed the $\sim t^2$ behavior is killed starting from the shortest times
and therefore fore some combinations of activity and temperature one gets
for the dynamical regimes as a function of time window:
diffusive, super-difussive or even ballistic, diffusive again.
This would certainly have a consequence in the characterization 
of the relaxation, since the dynamical structure factor is expected to
be affected accordingly.

\subsubsection{Dynamical structure factor at fixed activity}

\begin{figure}[t!]
\centering
\includegraphics[width=0.43\textwidth]{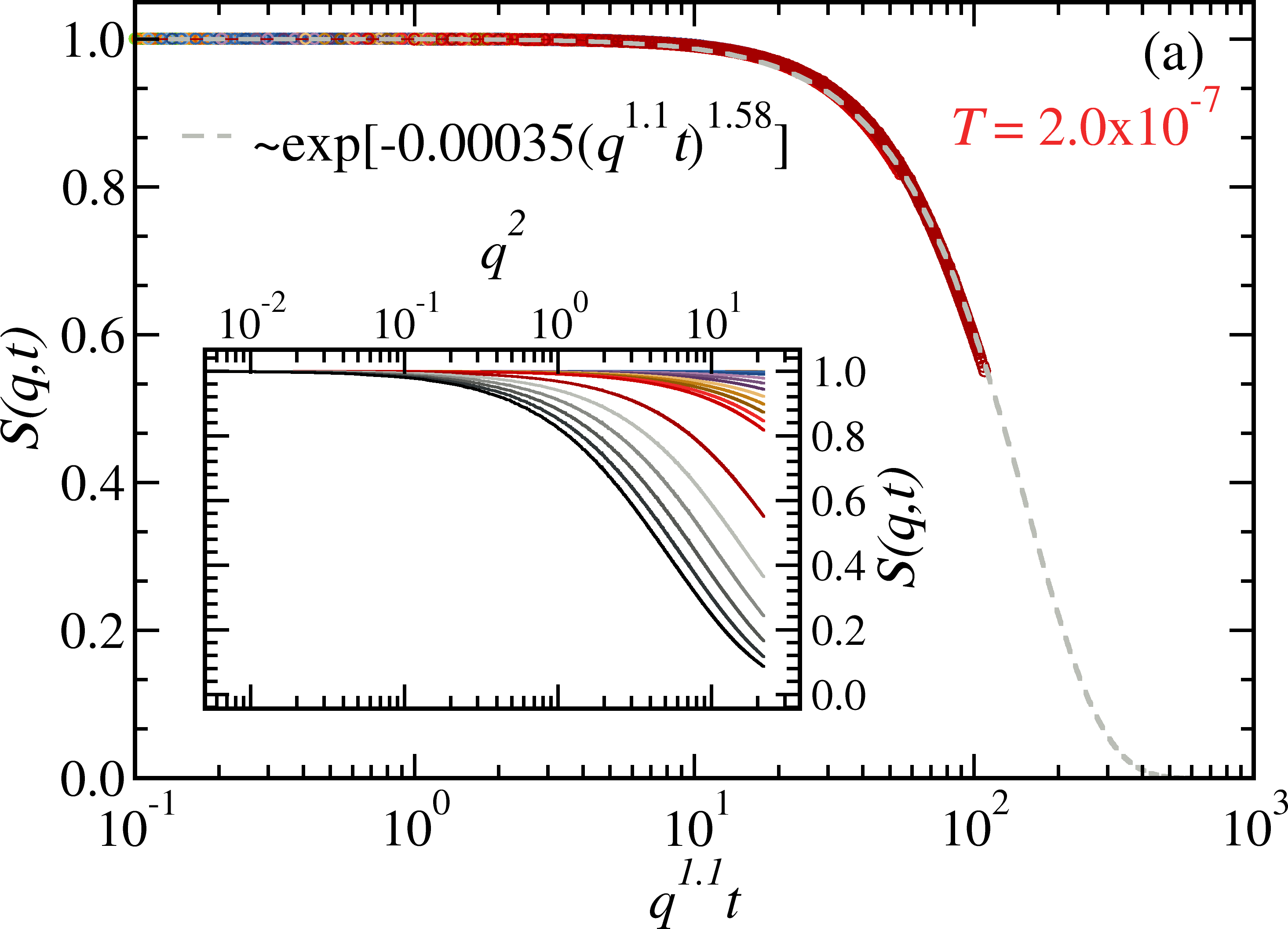} 
\includegraphics[width=0.43\textwidth]{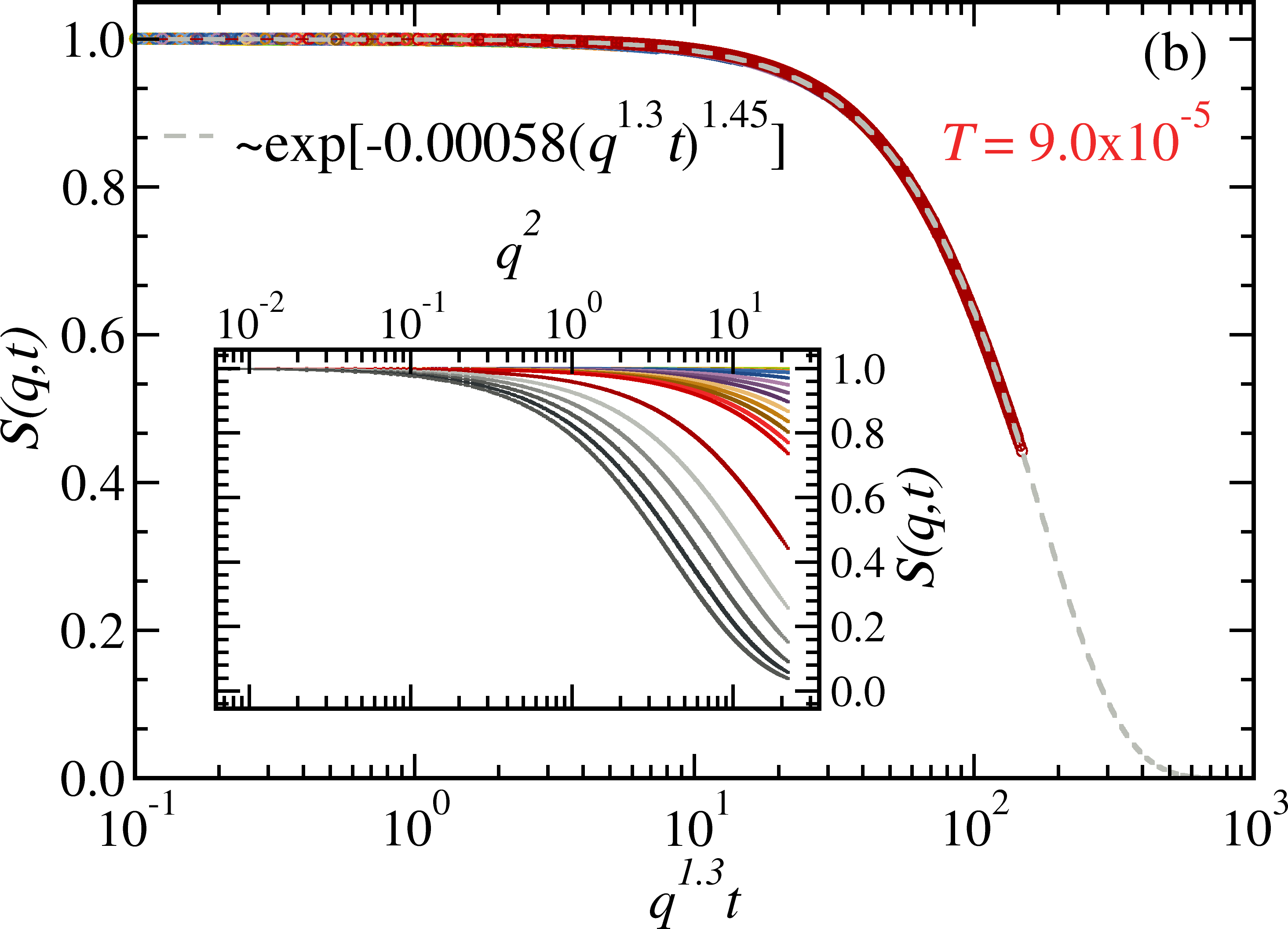} 
\includegraphics[width=0.43\textwidth]{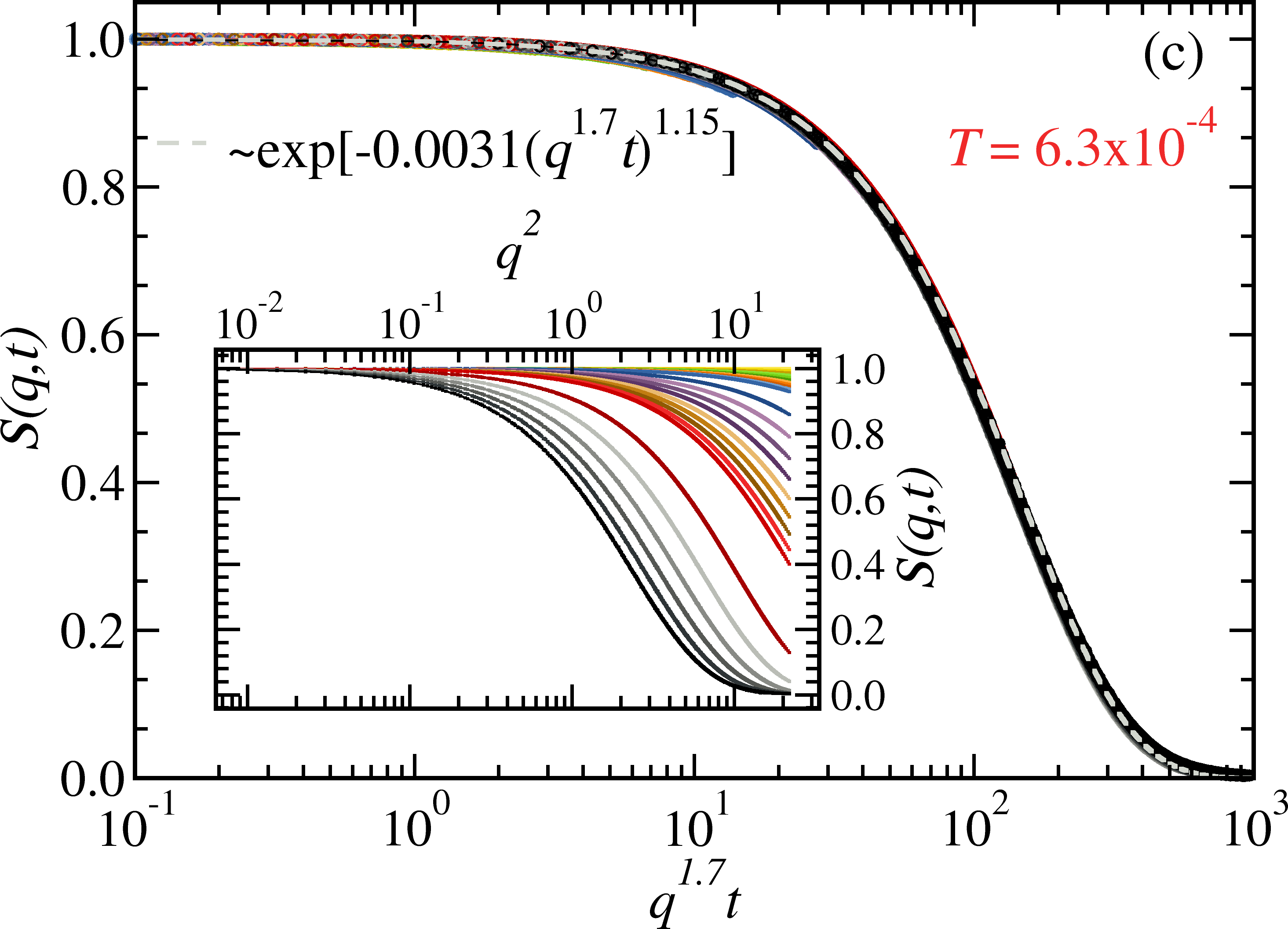} 
\caption{ 
{\it Dynamical structure factor $S(q,t)$ relaxation at short times 
and different temperatures but fixed activity}.
In all panels, the insets shows $S(q,t)$ as a function of $q^2$ for time 
windows up to $t=70$.
The dark-red curve corresponds to duration of plastic events, 
$\tau_{\tt ev} = 20$. $\epsilon_{0} = 0.1$. $L=32$.  
\textbf{(a)} The mainplot collapses the curves corresponding to 
time-windows $t \leq \tau_{\tt ev}$ by plotting $S(q,t)$ as function 
of $q^{1.1}t$.
The dashed line shows a compressed exponential of $(q^{1.1}t)$ with $\beta\simeq 1.58$.
\textbf{(b)} The mainplot collapses the curves corresponding to 
time-windows $t \leq \tau_{\tt ev}$ by plotting $S(q,t)$ as function 
of $q^{1.3}t$.
The dashed line shows a compressed exponential of $(q^{1.3}t)$ with $\beta\simeq 1.45$.
\textbf{(c)} The mainplot collapses the curves corresponding to 
all time-windows by plotting $S(q,t)$ as function 
of $q^{1.7}t$.
The dashed line shows a compressed exponential of $(q^{1.7}t)$ with $\beta\simeq 1.15$.
}
\label{fig:Sq_ThermalFixActivity}
\end{figure}

Thermal agitation on the particle tracers interferes in their persistent movement
at short times and this has also consequences on $S(q,t)$.
Figure~\ref{fig:Sq_ThermalFixActivity} shows the dynamical structure factor $S(q,t)$ 
in the short-time regime 
for three different temperatures ($T=2.0\times10^{-7}$, $9.0\times10^{-5}$ and 
$6.3\times10^{-4}$) at fixed activity, i.e., corresponding to the two extremes 
and an intermediate curve of Fig.~\ref{fig:msd-thermal-fixed-ratio}.
In the raw data of the insets one can appreciate that for a given time-window
(same curve color in the three panels) and same $q$, $S(q,t)$ has relaxed more
at higher temperatures.
This is intuitive since thermal agitation contributes to the decorrelation from
the initial configuration.
Nevertheless, the interesting characteristics lies in the functional form 
of the relaxation.
For each temperature we seek for structure factor behavior of the form 
$S(q,t) \propto \exp\left[-A (q^{\alpha}t)^{\beta}\right]$.
In panels (a) and (b) we have collapsed curves corresponding to time-windows
up to the duration of the plastic event ($t\lesssim \tau_{\tt ev}$),
while in (c) the good collapse with a single $\alpha$ extends to times above
$\tau_{\tt ev}$.
We observe that both $\alpha$ and $\beta$ varies with temperature.
The results for the lower temperature here (panel (a)) are consistent
with the athermal case presented in Fig.~\ref{fig:Sq_athermal_ballistic}.
As temperature increases, $\alpha$ increases from $\sim 1.1$ to $\sim 1.7$
while $\beta$ decreases from $\sim 1.58$ to $\sim 1.15$.

This goes in the direction of a crossover between a persistent movement
regime and a diffusive regime, as postulated in 
Fig.~\ref{fig:msd-thermal-fixed-ratio}.
Indeed, if we take the mean-field exponents~\cite{Bouchaud-InBook2008} 
as a guide, we may expect $\{\alpha,\beta\}$ moving from $\{1,3/2\}$ to 
$\{2,1\}$ as the temperature increases and the tracers at short times go
from ballistic to diffusive.
Yet, all exponents estimated from data are {\it effective} and not exactly
matching the mean-field ones.
One can argue that this is caused on one hand due to the crossover between
dynamical regimes at $t \sim \tau_{\tt ev}$ itself, but more importantly,
due to the fact that the system is in finite dimensions and non-trivial
correlations among plastic events are expected to play a role.

\begin{figure}[t!]
\includegraphics[width=\columnwidth]{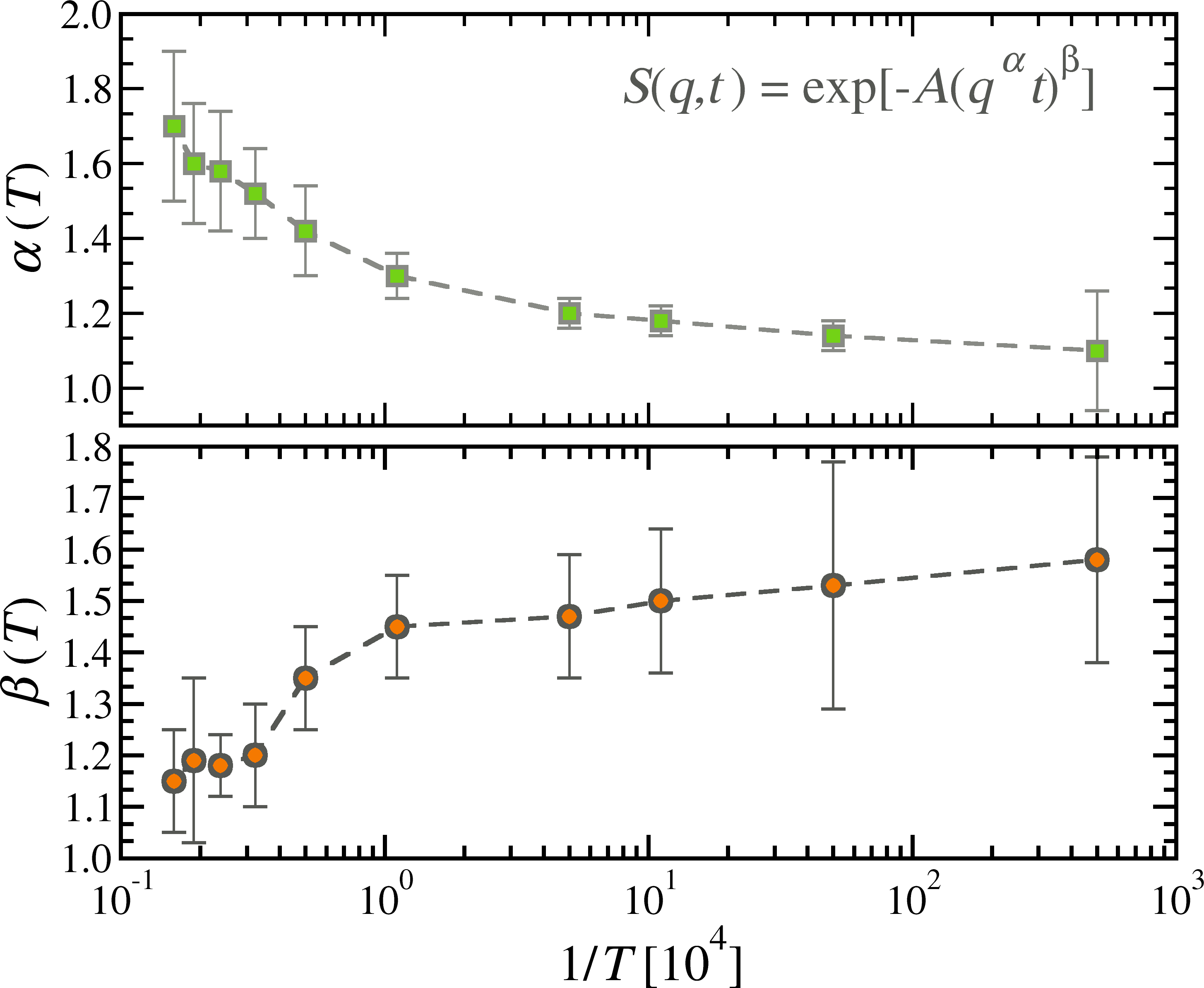}
\caption{\label{fig:relaxation_exponent} 
{\it Relaxation exponent $\alpha$ and $\beta$ at short times for different 
temperatures but fixed activity.} 
The upper-panel shows $\alpha$ vs. $1/T$ and the lower-panel displays 
$\beta$ vs. $1/T$.
The $1/T$ axis is displayed in log-scale simply for a better visualization.
The estimates correspond to collapses and fits as those of Fig.~\ref{fig:Sq_ThermalFixActivity}.
$\tau_{\tt ev}$ = 20.0. $\epsilon_{0}$ = 0.1. $L = 32$.
}
\end{figure}

Figure~\ref{fig:relaxation_exponent} shows the evolution of $\alpha$ and $\beta$
with temperature.
The exponents have been measured for several temperatures from collapses and 
fits as the ones in Fig.~\ref{fig:Sq_ThermalFixActivity}.
The error bars are estimated from independent fit approaches and then doubled.
Let us notice first in the bottom panel how $\beta$ increases
from $\sim 1.1$ to $\sim 1.6$ as the temperature is lowered. 
These results can be compared with the recent reports in
Zr$_{46.8}$Ti$_{8.2}$Cu$_{7.5}$Ni$_{10}$Be$_{27.5}$ metallic glass 
($T_g \simeq 596 K$) presented in Ref.~\cite{AminiPRM2021}, where 
$\beta$ decreases with increasing temperature smoothly from 
$1.75\pm 0.14$ at 523~K to $0.67\pm 0.07$ at 618~K.
A stretched exponent $\beta<1$ indicates that the sample has 
reached the super-cooled liquid state (inaccessible in our model).
Earlier~\cite{RutaPRL2012}, the shape factor $\beta$ was studied
for Mg$_{65}$Cu$_{25}$Y$_{10}$ ($T_g \simeq 405 K$), showing that it
remained in the range 1.7-1.4 before plunging when approaching $T_g$.
Our results for the temperature dependence of $\beta$ are also 
consistent with what has been reported in simulations of 
network-forming gels~\cite{Gado2017}.
Accompanying the gradual change of $\beta$ with temperature, the 
exponent $\alpha$, that defines how the characteristic relaxation
time depends on the wave vector, decreases with decreasing temperature, 
from $\sim 1.7$ (diffusive-like) to $\sim 1.1$ (ballistic-like).
In other words, for a fixed $q$ the characteristic relaxation time 
$\tau\sim q^{-\alpha}$ decreases with increasing temperature.
Again, in agreement with~\cite{RutaPRL2012,Gado2017,AminiPRM2021}. 

We therefore propose that the shape exponent variation with temperature 
in measured relaxation of amorphous solids can be understood by the
competition of rather persistent displacements on its essential constituents,
induced by the occurrence of plastic events nearby, on one hand, and the 
intrinsic thermal agitation at which they are subject that tends to generate 
a Browninan motion of particles.

\subsubsection{Displacements distributions with thermal agitation}

\begin{figure}[t!]
\includegraphics[width=\columnwidth]{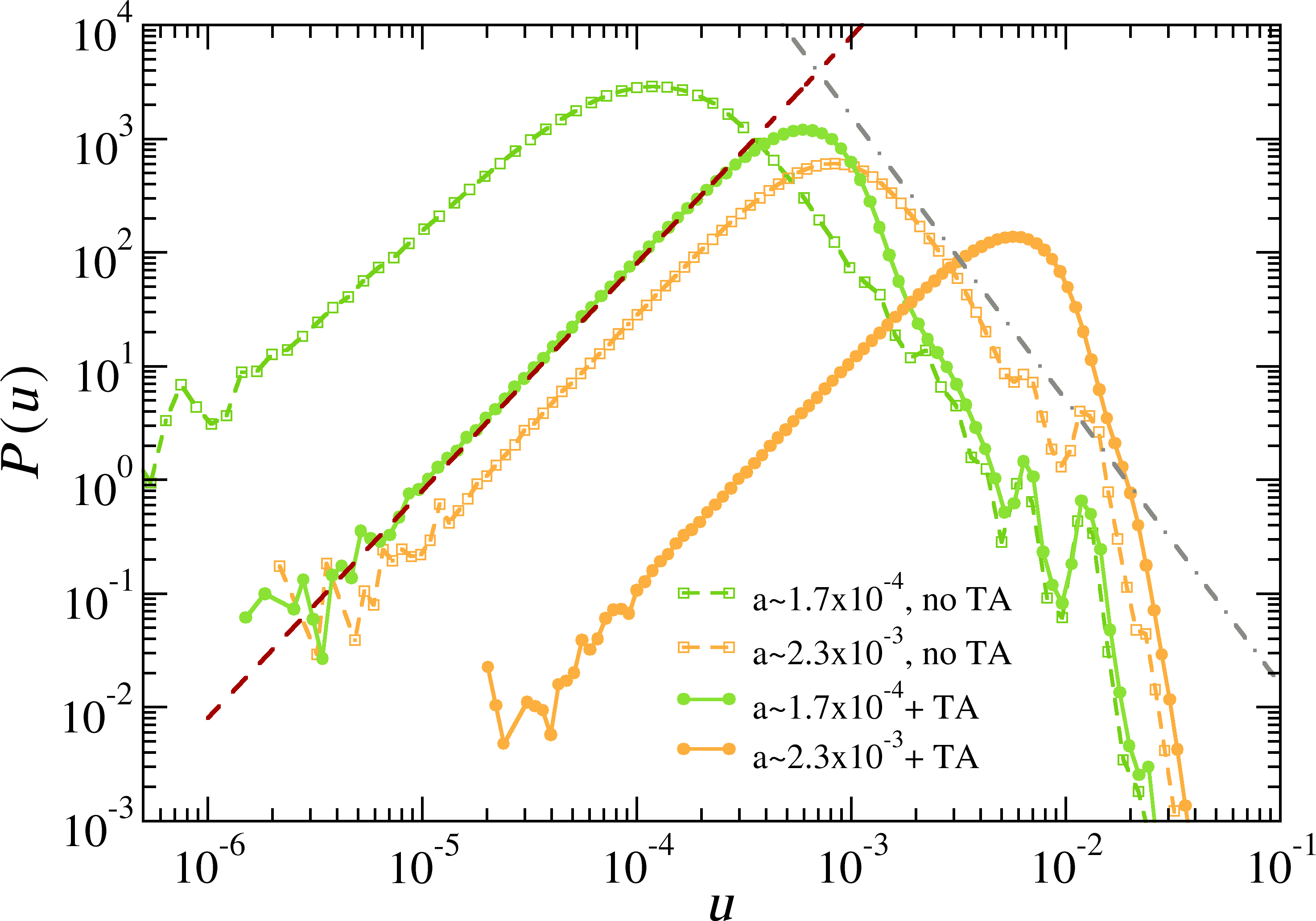}
\caption{\label{fig:Pofu_thermal} 
{\it Displacement distribution $P(u)$ for different mean activities
with and without thermal agitation (TA)}.
Green curves relate to an activity $\left<a\right> \simeq 0.00017$: 
while open symbols repeat the data for $T=0.035$ in Fig.~\ref{fig:Pofu_athermal}, 
closed symbols correspond 
to $T\simeq 8\times10^{-7}$ ($B$ changes to maintain activity) 
an thermal agitation proportional to $T$ has been included
in the tracers.
Orange curves relate to $\left<a\right> \simeq 0.0023$
($T=0.05$ in Fig.~\ref{fig:Pofu_athermal}), and the thermal 
agitation included in this case is $T\simeq 8\times10^{-5}$. 
Red dashed-line, gray dashed-point line  
show $P(u)\sim u^{2}$ and $P(u)\sim u^{-5/2}$, respectively.
The tracers displacements absolute values $u$ are measured each 
$0.1$ time units.
$\tau_{\tt ev} = 1.5$. $\epsilon_{0} = 1.0$. $L = 32$.
}
\end{figure}

To complete the picture of what happens when we include thermal agitation
in the tracer particles, we present here the displacement distributions
$P(u)$ for a couple of temperatures with the addition of thermal agitation 
on the tracer movement.
Figure\ref{fig:Pofu_thermal} shows $P(u)$ for a couple of
different temperatures to see the thermal agitation effect
on the displacements.
Green solid points correspond to a relatively low temperature of 
$T\simeq 8\times10^{-7}$ (and $B=2.3\times 10^{-5}$ is equally 
low such that $\left<a\right> \simeq 0.00017$)
We can see that, although the large $u$ tail of P(u) is
still dominated by the plastic activity induced displacements,
the thermal agitation or Brownian motion of the tracers start to 
dominate the small displacements, clearly shifting the $P(u)$ peak 
position
(comparison is done with the \{$T=0.035$, $B=1$\} curve of
Fig.~\ref{fig:Pofu_athermal}).
For the orange solid points $T\simeq 8\times10^{-5}$ is already large
enough to completely erase the hallmark of the displacements of
mechanical origin
(comparison is done with the \{$T=0.05$, $B=1$\} curve of
Fig.~\ref{fig:Pofu_athermal}).
The $P(u)\sim u^{-5/2}$ behavior at large $u$ completely disappears,
while the $P(u)\sim u^{2}$ scaling at small $u$ still shows,
but now directly generated by a genuine Brownian motion 
instead of being a result of an incoherent superposition 
of displacements with mechanical origin
(see App.\ref{app:Pu}).
The disappearance of the $P(u)\sim u^{-5/2}$ tail at large enough
temperatures, justifies a diffusive regime $S(q,t)\sim \exp[-Aq^2t]$
stepping in sooner than in the pure plastic activity case, as we verify.

\subsection{Reactivation-times and dynamical heterogeneity}

\begin{figure}[t!]
\includegraphics[width=\columnwidth]{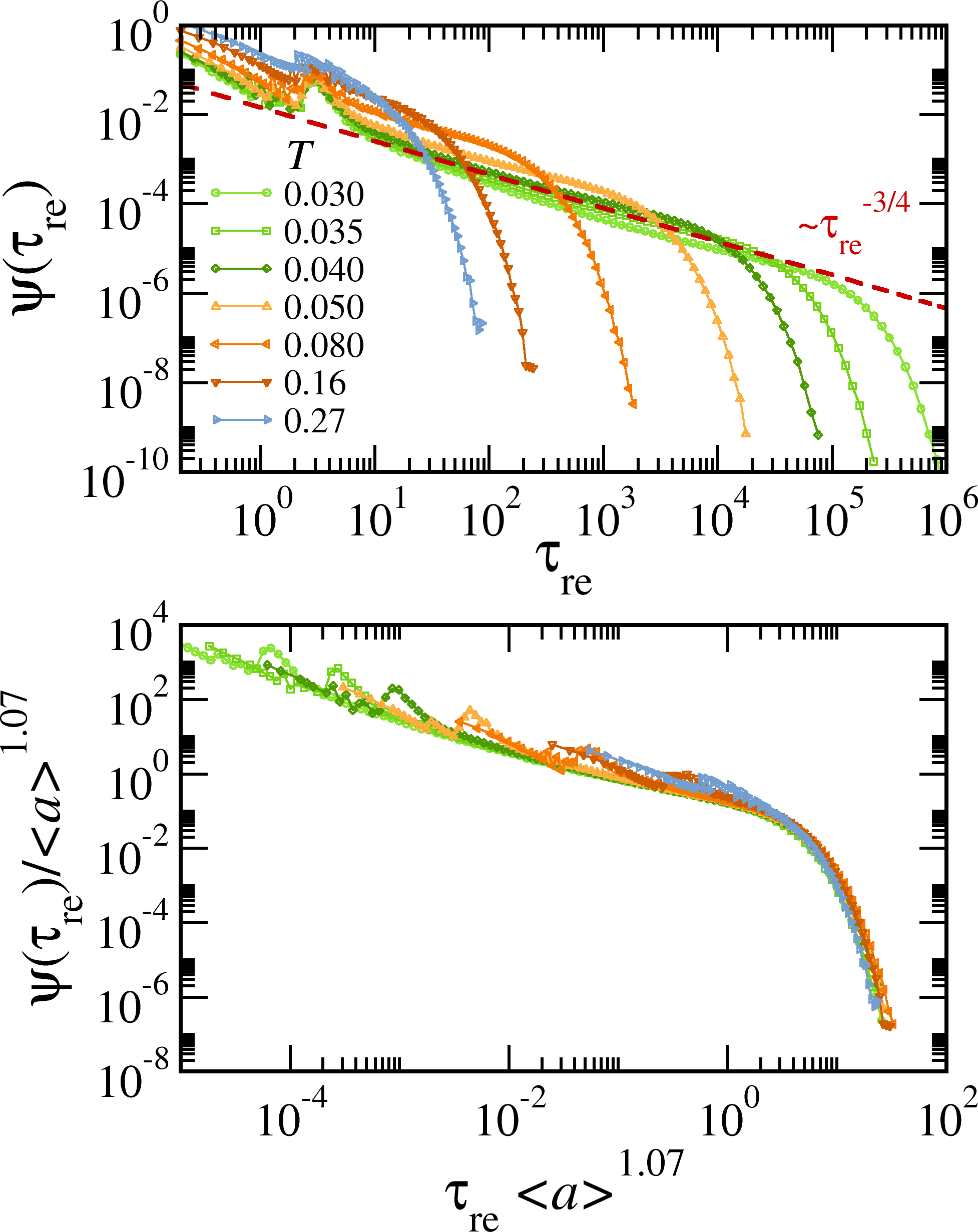}
\caption{\label{fig:PofTre} 
{\it Distribution of reactivation times $\Psi(\tau_{\tt re})$ for different temperatures.} 
$\tau_{\tt re}$ are the time lapses between consecutive occurrences of a plastic event 
in the same system patch.
Upper-panel shows raw data for different temperatures (activity levels).
Red-dashed line displays a power-law $\sim \tau_{\tt re}^{-\omega}$, with the 
exponent $\omega=3/4$.
Lower-panel rescales the curves plotting 
$\Psi(\tau_{\tt re})/\left<a\right>^{1.07}$ vs $\tau_{\tt re}\left<a\right>^{1.07}$. $B = 1.0$.
$\tau_{\tt ev}$ = 1.5. $\epsilon_{0}$ = 1.0. $L=32$, 
}
\end{figure}

So far we have concentrated in quantities that are computed from
the tracer particle movements. 
We now focus on the evolution of the elasto-plastic system itself and
jump to the analysis of the intermittent and heterogeneous dynamics 
observed in the quiescent amorphous solid.
System patches locally yield, either activated by temperature or 
over-stressed by the action of other sites; 
then they recover elasticity (on average) after a time $\tau_{\tt ev}$.
The time takes to see two consecutive activations of a given site will 
depend somehow trivially on the global activity level, but each occurence 
can be difficult to predict.
In fact, this quantity -that we call ``reactivation-time'' 
$\tau_{\tt re}$- is broadly distributed, in particular
at low temperatures.

In Fig.\ref{fig:PofTre} we show $\Psi(\tau_{\tt re})$ for different temperatures.
Beyond short $\tau_{\tt re}$, one sees a power-law distribution 
$\Psi(\tau_{\tt re}) \sim \tau_{\tt re}^{-\omega}$, with $\omega\simeq 3/4$,
followed by an exponential upper cutoff.
Interestingly, the cutoff depends on temperature and moves to larger and larger 
reactivation times as $T$ is decreased.
For our lowest temperatures, $\Psi(\tau_{\tt re})$ expands over almost 7 orders 
of magnitude.
As expected, $\tau_{\tt re}$ seems be proportional to an activation rate 
that scales as the inverse of activity $\left<a\right>^{-1}$, and we can 
use that to collapse the distribution tails in Fig.\ref{fig:PofTre}(bottom).
We find that an exponent close but not exactly unity makes the best collapse,
happening when we plot $\Psi(\tau_{\tt re})/\left<a\right>^{1.07}$ vs 
$\tau_{\tt re}\left<a\right>^{1.07}$, preserving the normalization of the 
distributions rather than the power-law exponent.
As already discussed in~\cite{FerreroPRL2014}, the interaction among plastic events
cause the emergence of a characteristic short reactivation time of the order of 
$\tau_{\tt ev}$, presumably due to neighbor sites alternately triggering each other
during a burst of activity.
That is the little peak observed in all curves, that looks sharper at low temperatures.
At the same time, correlations among sites induce a fat-tailed distribution of 
reactivation times, that are increasing as $T$ decreases, which otherwise would decay 
exponentially in a Poissonian way.
At low temperatures, the intermittency of the plastic events at a given site is 
determined by events happening in other regions of the system, it's intrinsically 
a spatial property.
Furthermore, notice that the exponent $\omega$ controlling the power-law regime
depends on dimension: 
It was $\omega \simeq 2/3$ in $d=2$ and it's now $\omega\simeq 3/4$ in $d=3$.
Although one can be tempted to think on a $d/(d+1)$ dependence, we don't 
have an argument for this observation.
Intermittency was argued as a possible explanation of the shape 
exponent $q$-dependence~\cite{DuriCi-EPL2006}.
While we don't address that problem in the present work, it's worth mentioning that
in EP models the activation times of a block is a direct observable.
So this is advantageous for studies of intermittency of plastic activity.

\begin{figure}[t!]
\includegraphics[width=\columnwidth]{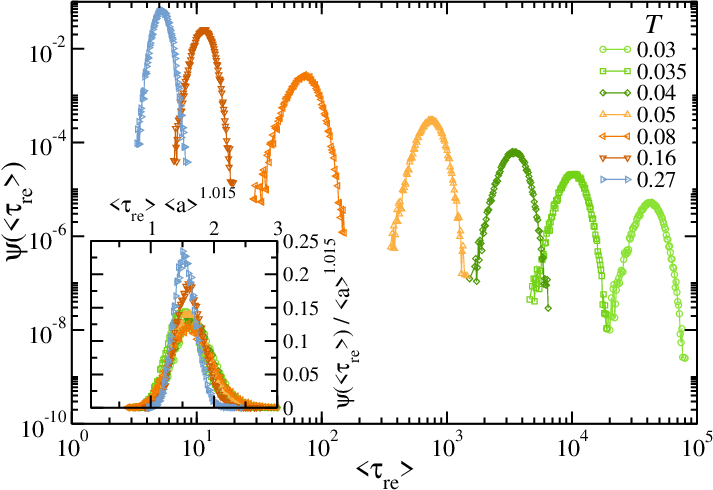}
\caption{\label{fig:Pof_meanTre} 
{\it Distribution of mean reactivation times $\Psi(\left<\tau_{\tt re}\right>)$ 
for different temperatures.} 
$\left<\tau_{\tt re}\right>$ are the average time lapses between consecutive 
occurrences of a plastic event in each block of the system.
Averages are taken over 10$^{2}$ local measurements of $\tau_{\tt re}$ very well 
separated in time (non-correlated), and the histogram is computed with the 
statistics of $2^{15}$ blocks. $B = 1.0$. 
$\tau_{\tt ev}$ = 1.5. $\epsilon_{0}$ = 1.0. $L=32$.
}
\end{figure}

One is tempted to relate this broad distribution of reactivation times
to {\it dynamical heterogeneities}.
In fact, dynamical heterogeneities have been recently discussed in thermal EPMs. 
In~\cite{OzawaPRL2023}, it is argued that the patterns formed by the local 
persistence at different temperatures show that the dynamics is spatially 
heterogeneous over lengths that increase when lowering the temperature.
Nevertheless, at least in our case, it can be seen clearly that such 
heterogeneity in space is only momentary
and do not persist when the dynamics is integrated in time.
In other words, one does not find the paradigmatic dynamical heterogeneity 
of glasses where for example, as in spin glasses, 
persisting {\it fast} and {\it slow} regions can be 
identified~\cite{Ricci-TersenghiPRE2000,RomaPRL2006,FerreroPRE2012}.
In the absence of anisotropic fields or disorder in the local activation dynamics,
such as a quenched disorder or a dynamical disorder creating diverging correlation times,
the system is intrinsically {\it homogeneous} and we observe such homogeneity in the
time averages of $\tau_{\tt re}$.
In Fig.~\ref{fig:Pof_meanTre} we take the average reactivation time on each block 
during a long run, with care to average over uncorrelated samples, this is, taking 
well separated configurations in time, and we plot its distribution 
using our $2^{15}$ blocks as a representative statistics.
It is interesting to see that the mean value $\left<\tau_{\tt re}\right>$ is 
well peaked around a typical value that depends on temperature (activity).  
The moving peak/bump at large $\left<\tau_{\tt re}\right>$ is simply 
a signature of blocks needing to wait more and more to get re-activated 
as $T$ decreases.


\section{Discussion}
\label{sec:discussion}
\vspace{-0.3cm}

We have used a three-dimensional thermal elasto-plastic model 
with the addition of probe particles to follow the displacement 
fields and analyze the equilibrium structural relaxation dynamics of quiescent 
amorphous materials at finite temperatures. 
The results show that for sufficiently short times and when the
plastic activity is relevant, there is always a super-diffusive regime 
in the mean square displacement of the probe particles, after which 
they enter a crossover towards a diffusive behavior. 
The crossover is dominated by the typical duration of plastic events in agreement 
with the interpretation that the close to linear motion of the particles in the 
super-diffusive part is an elasticity-mediated phenomenon.

In addition, we observed a compressed exponential relaxation 
(with a shape exponent $\beta$ reminiscent of experimental results) 
in the dynamical structure factor $S(q,t)$ at short times, 
associated with the elasticity-mediated super-diffusive regime, 
that exists even in the equilibrium dynamics of this model. 
For a sustained plastic activity level, the $\beta\!>\!1$ exponent 
varies with temperature, decreasing as the thermal agitation of the 
particles becomes more and more relevant, eventually turning 
the compressed- into a simple-exponential ($\beta \sim 1$).

At long times, in the diffusive regime, the relaxation is always 
exponential. 
Furthermore, we observe the crossover from a $(q^{3/2}t)$ to a $(q^2t)$
diffusive relaxation when activity increases, as predicted by 
mean-field arguments~\cite{BouchaudPi-EPJE2001,BouchaudPi-EPJE2002}.
The displacements distribution $P(u)$ helps to interpret 
the behavior of $S(q,t)$, with the characteristic 
$P(u)\sim u^{-5/2}$ decay of large displacements 
(also expected from mean-field arguments) disappearing as
the mean activity increases.

We noticed that our approach does not reproduce in any regime 
the characteristic stretched exponential relaxation of 
glasses~\footnote{The intermediate sub-diffusive, `statistical caging'
and associated stretched relaxation regime observed by some of 
us in ref.~\cite{FerreroPRL2014} turned out to be a feature of 2D systems only.}.
Reproducing the commonly observed stretched exponential relaxations at long times seems 
not possible without further complexification of the EP model under consideration.
A common belief is that stretched exponentials are created by the 
superposition of a large distribution of relaxation times in the 
glassy heterogeneous dynamics~\cite{PalmerPRL1984}.
In our thermal elasto-plastic model, we show indeed a feature of dynamical 
heterogeneity (similar to~\cite{OzawaPRL2023}): 
a very wide distribution of activation `waiting' times is observed 
instantaneously. 
Yet, the system remains homogeneous in the time-integrated dynamics, 
and stretched relaxation never happens. 
Without a quenched disorder or other kind of imposed persistent heterogeneity, 
we don't believe that stretched exponentials can be reproduced in elasto-plastic models. 

Still, other avenues to explore modifications of elasto-plastic models 
towards the characterization of glasses are possible.
Recent studies have shown that stretched exponential behavior in glasses 
could be observed already on a local scale relevant to the localized 
relaxation events~\cite{shang2019local}, and stretched exponentials 
can thus be obtained independently of a broad distribution of relaxation times. 
Also, stretched exponential relaxation can be obtained by appropriately 
tweaking the dynamics of local relaxation events and weighting their 
interactions, at least in a mean-field approach~\cite{TrachenkoJPCM2021}.
We can envision that modifications on how local stress relaxation occurs 
in plastic events in elasto-plastic models, e.g. abandoning the simplistic 
instantaneous, simple exponential or linear options explored so far, 
might also allow for a stretched relaxation regime to arise in the dynamics,
irrespective of the system homogeneity.
Furthermore, EP models could also serve as a numerical tool 
to compare with recent under-pressure relaxation measurements in 
BMGs~\cite{CornetAM2023}, through an easy modification of the model implementation and parameters.

Finally, we expect that our analysis of the interplay between plastic activity induced
displacements and simple agitation due to finite temperature would inspire
ongoing experiments on systems whose elementary constituents and are sensitive 
to Brownian motion to try to discriminate both contributions in the 
auto-correlation functions of the scattered intensity;
for example, by independently assessing the `mean plastic activity' during the
experiment as temperature is varied.

\begin{acknowledgments} 
We acknowledge support from the CNRS IRP Project ``Statistical Physics of Materials''
and PIP 2021-2023 CONICET Project Nº 11220200100757CO.
EEF acknowledges support from the Maria Zambrano program of the 
Spanish Ministry of Universities through the University of Barcelona,
and MCIN/AEI support through PID2019-106290GB-C22.
\end{acknowledgments}

\appendix

\section{Steady-state stress distribution}\label{app:stress_distributions}

This Appendix shows the stationary distribution of stresses
in our elasto-plastic model at different temperatures or activities.
In the way the model is defined, it only makes physical sense if the 
plastic activity is ``low enough''.
High activity levels, induce a notable fraction of the blocks to 
remain over-stressed and ``out of the box'' $\left[-\sigma_y, \sigma_y\right]$.
That is to be avoided if we want to make sense of the results.
Therefore, our control observable has been the distribution of stress 
at each given temperature or mean activity.

\begin{figure}[t!]
\includegraphics[width=\columnwidth]{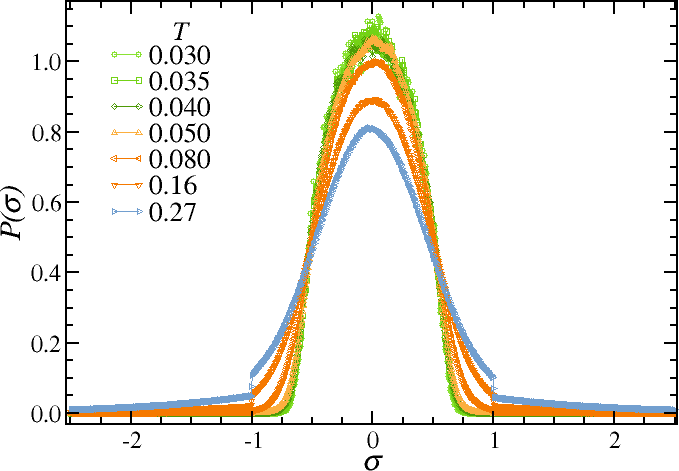}
\caption{\label{fig:stress_athermal_tracers} 
Stress distribution for different temperatures in the stationary state.
Parameters: $B=1$ and $T$ ranging from $0.03$ to $0.27$ as indicated in the 
legends, with respective mean activities $\left<a\right>\simeq 4.5\times10^{-5}$, $1.7\times10^{-4}$, 
$5\times10^{-4}$, $2.3\times10^{-3}$, $0.023$, $0.14$, $0.3$. 
$\tau_{\tt ev}$ = 1.5. $\epsilon_{0} =$ 1.0. $L = 32$.
}
\end{figure}

Figure ~\ref{fig:stress_athermal_tracers} shows the normalized stress 
histograms $P(\sigma)$ corresponding to the steady states used to produce Fig.~\ref{fig:msd_athermal_tracers}.   
As plastic activity increases, the stress distributions become wider and wider.
The local yield stresses, being set to $\sigma_y = \pm 1$, make the distribution 
to show a pseudo discontinuity as the stress $\sigma$ goes out of the box.
Let us recall than for these steady states, the prefactor of the Arrhenius term 
was simply $B=1$ and $T=0.03,...,0.27$ induced mean activities of 
$\left<a\right> \simeq 4.5\times10^{-5},..., 0.3$.

\begin{figure}[b!]
\includegraphics[width=\columnwidth]{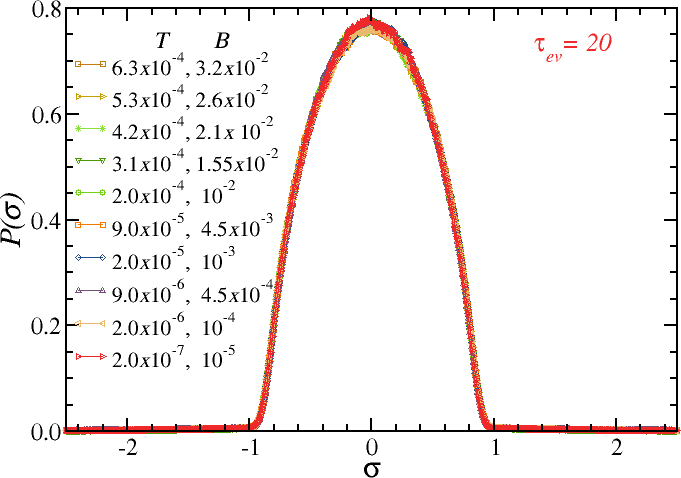}
\caption{\label{fig:stress_thermal_tracers} 
Stress distribution for different temperatures in the stationary state. 
Parameters $B$ and $T$ as shown in the legends, preserving a mean 
activity $\left<a\right> \simeq 0.05$.
$\tau_{\tt ev}$ = 20, $\epsilon_{0}$ = 0.1. $L = 32$.
}
\end{figure}

Moving to the analysis of the interplay between plastic activity and thermal agitation we have fixed 
a plastic activity to $\left<a\right> \simeq 0.05$ 
varying accordingly $B$ and $T$.
Figure~\ref{fig:msd-thermal-fixed-ratio} shows the distributions $P(\sigma)$ 
for the steady states of such cases.
As one expects, the stress distribution only depends on the mean activity and 
all curves coincide.

\section{Mean plastic activity $\left<a\right>$ vs temperature}\label{app:activity_vs_T}
The mean plastic activity $\left< a \right>$ is directly 
controlled by the temperature $T$ and the parameter $B$ through Eq.~\ref{eq:activationrulesEPM}, but it also depends on the effect 
that a plastic event has on other sites, which is modulated by 
$\epsilon_0$, and in the plastic events duration $\tau_{\tt ev}$.
In this Appendix, we show activity as a function of temperature 
for the case corresponding to the steady states of 
\{$B=1$, $\tau_{\tt ev}=1.5$, $\epsilon_0=1.0$\} and 
\{$B=0.001$,  $\tau_{\tt ev}=20$, $\epsilon_0=0.1$\}.

\begin{figure}[t!]
\includegraphics[width=0.95\columnwidth]{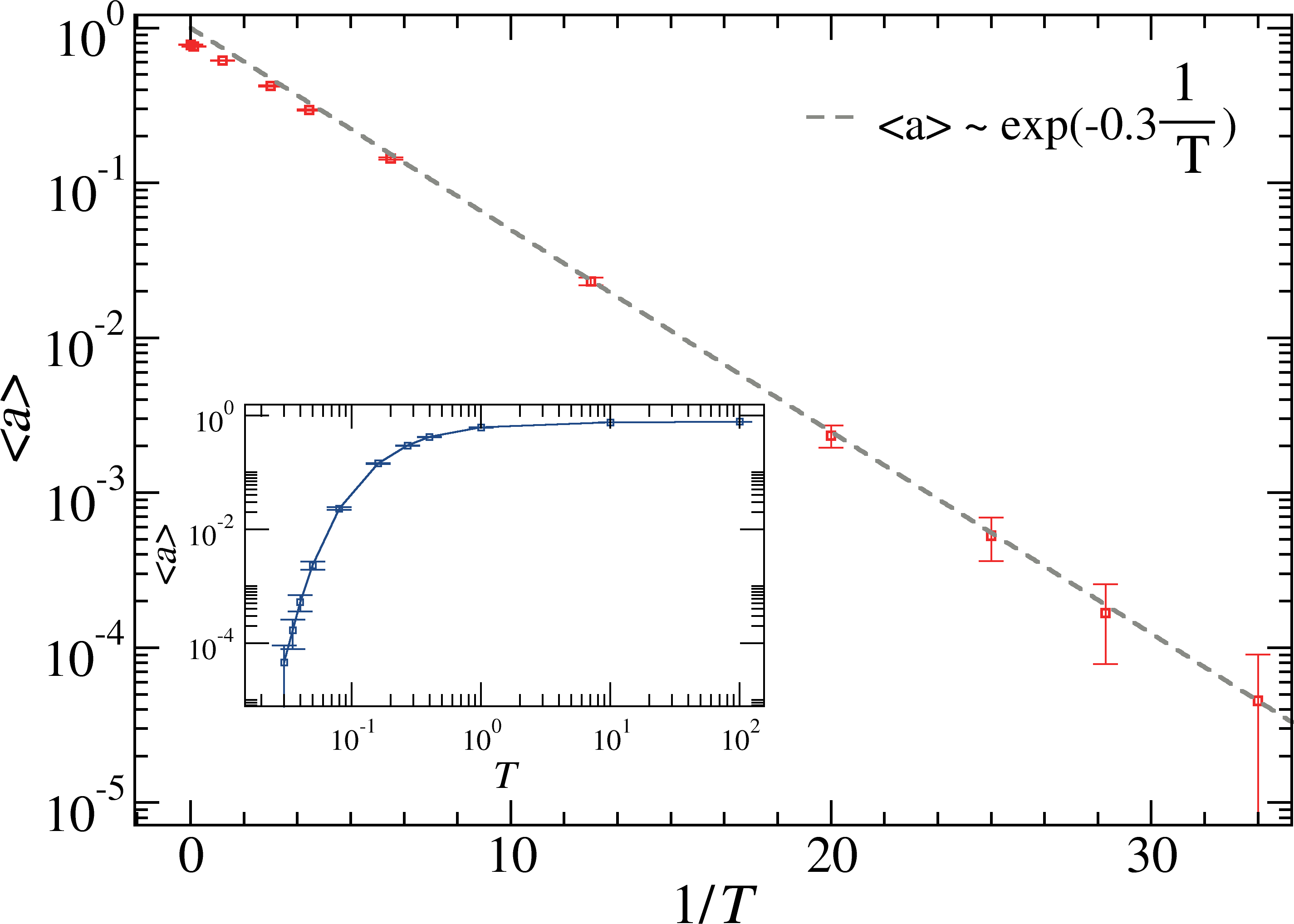}
\includegraphics[width=0.95\columnwidth]{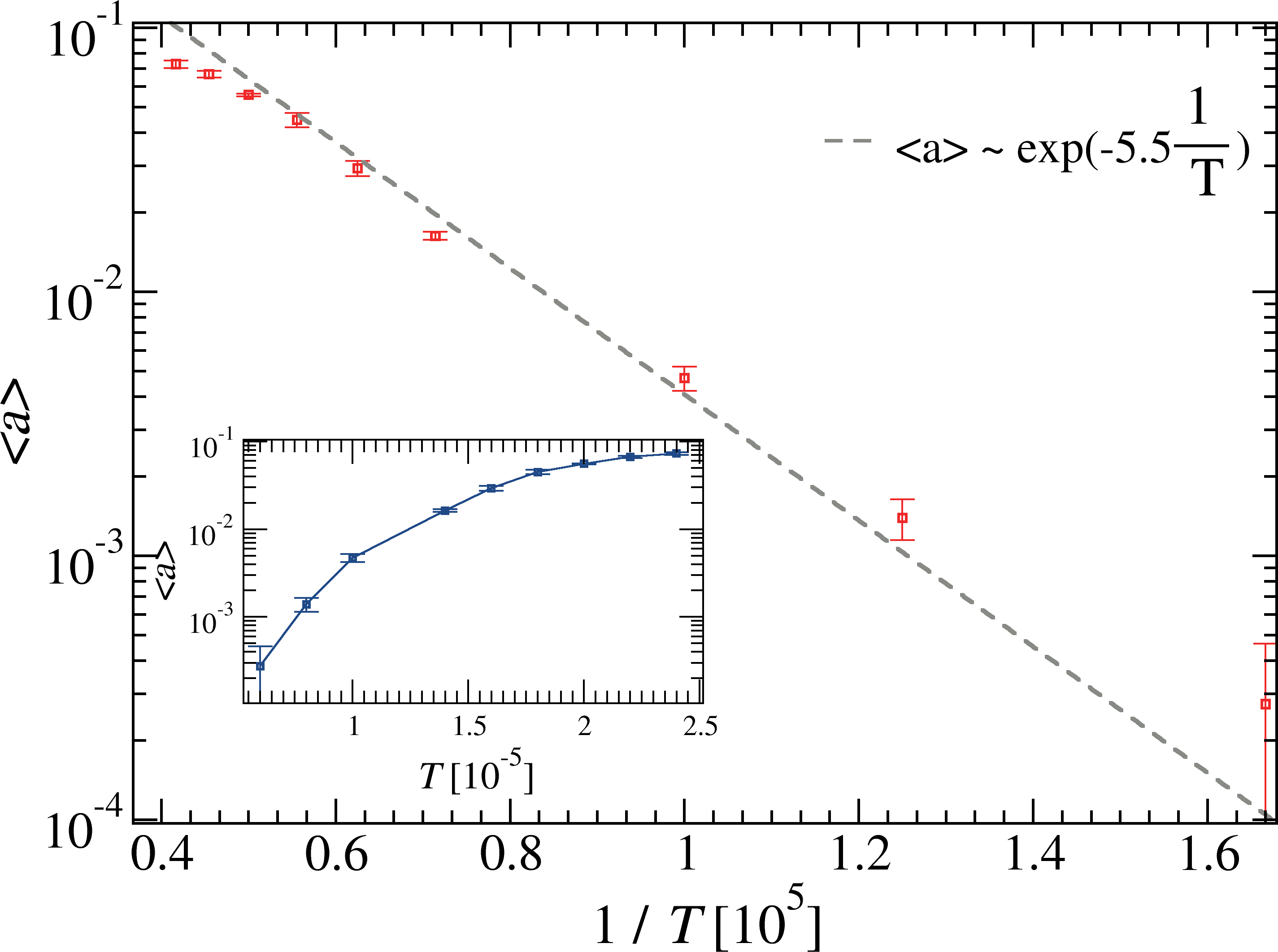}
\caption{\label{fig:activity_vs_T} 
Main plots: Mean plastic activity $\left< a \right>$ vs $1/T$ in the stationary state
Insets: $\left< a \right>$ vs $T$.
Upper panel: $B=1$, $\tau_{\tt ev} = 1.5$, $\epsilon_{0} = 1.0$ and $L = 32$.
Lower panel: $B=0.001$, $\tau_{\tt ev} = 20.0$, $\epsilon_{0} = 0.1$ and $L = 32$.
}
\end{figure}

Figure~\ref{fig:activity_vs_T} upper panel shows the mean activity $\left< a \right>$
as a function of $1/T$ for the same steady states as 
in Fig.~\ref{fig:msd_athermal_tracers} (plus some extra temperatures).
The behavior of the activity is exponential in $1/T$, as expected.
The inset, plotted in log-lin, shows $\left< a \right>$ vs $T$.
We can appreciate that for $T\to 0 $  the activity decays fast, while, 
on the other hand, it saturates at large temperatures. 
The lower panel shows $\left< a \right>$ vs. $1/T$ (main panel) and $T$ (inset)
for the same steady states as in Fig.~\ref{fig:msd-thermal-fixed-B}.

\section{Displacement distribution $P(u)$: different acquiring time
and Brownian motion role}\label{app:Pu}

This Appendix intends to help in the interpretation of the 
results for the displacements distribution $P(u)$.
On one hand, the definition of $u$ itself depends on a 
time window observation.
$u$ is defined as the displacement of a tracer in a given time 
$\Delta t$.
Changing $\Delta t$, changes $u$ and therefore also $P(u)$.
On the other hand, it's important to understand the limit cases. 
\\

{\it Displacement distribution for different definitions of $u$.} 

\begin{figure}[t!]
\includegraphics[width=\columnwidth]{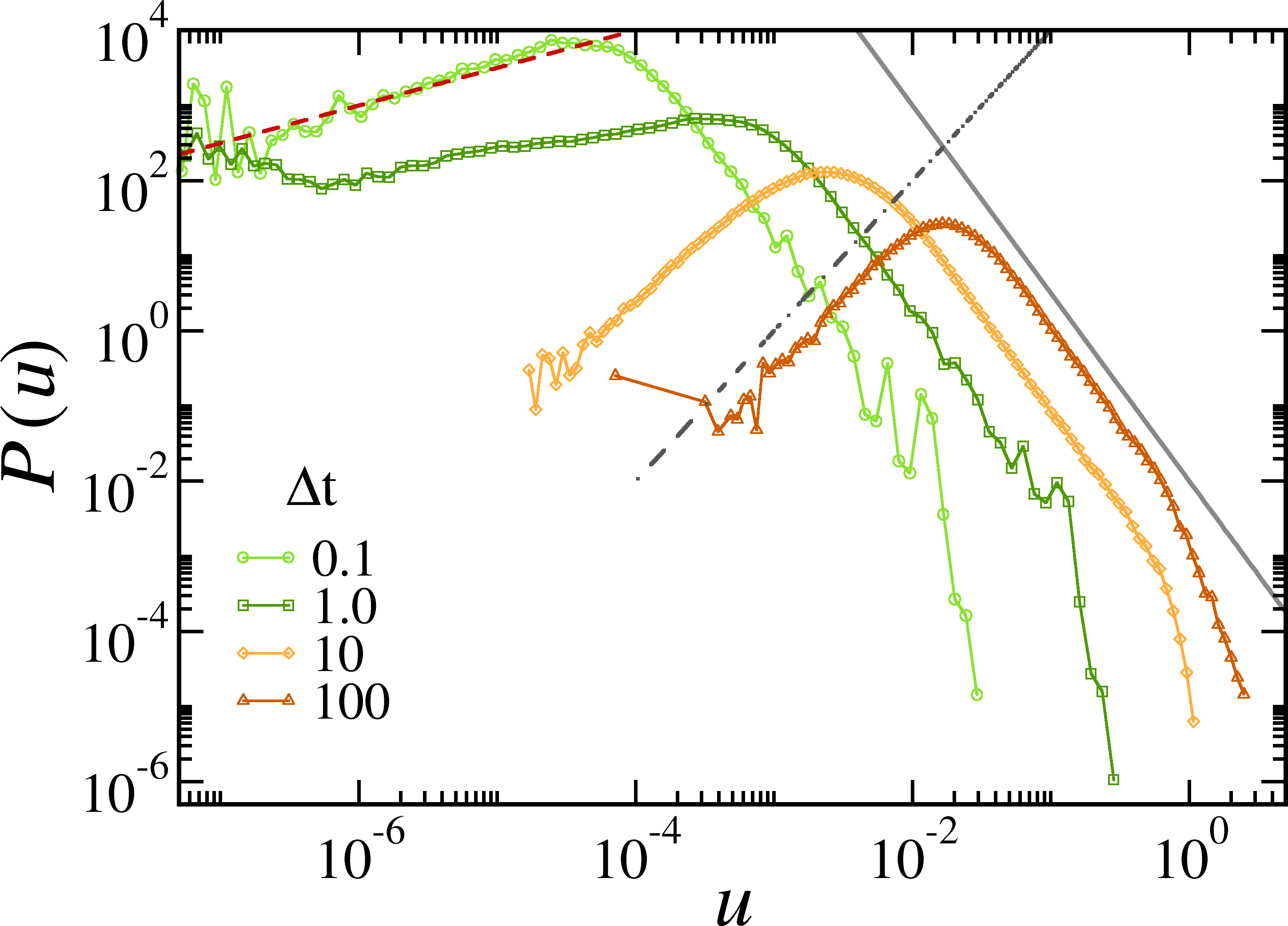}
\caption{\label{fig:Pu_different_dataAdquiring} 
Displacement distribution $P(u)$ where $u$ is defined as a tracers displacement 
time window $\Delta t$ (different values of $\Delta t$ indicated in the legends). 
Parameters: $T = 0.03$, $\tau_{\tt ev} = 1.5$, $\epsilon_{0} = 1.0$, $L = 32$. 
Red-dashed, gray point-dashed and gray full-line
show $P(u) \sim u$, $\sim u^{2}$ and $\sim u^{-5/2}$, respectively.} 
\end{figure}

Figure~\ref{fig:Pu_different_dataAdquiring} shows displacement 
distributions $P(u)$ for different definitions of the displacements $u$
at low plastic activity.
In particular, we define $u$ as the tracer displacement in a time windows
$\Delta t$.
We show data for four different definitions $\Delta t = 0.1, 1.0, 10.0$ 
and $100.0$. 
Notice that $\Delta t = 0.1$ is the case used in Figs.~\ref{fig:Pofu_athermal}
and~\ref{fig:Pofu_thermal}.
A favorable consequence of increasing $\Delta t$ is that the noisy peaks
of the $P(u)$ tail at large $u$ smooth-out.
On the other hand, the distribution shrink, the ranges of $u$ where
power-laws could be fitted become thinner and we may evan loose some information
about very small displacements.
Still, even the largest $\Delta t$ shows clearly the power law tail at large displacements $\sim u^{-5/2}$, 
and $\sim u^{2}$ for the smallest displacements 
observed within that definition.\\

{\it Displacement distribution of purely Brownian particles} 

\begin{figure}[t!]
\includegraphics[width=\columnwidth]{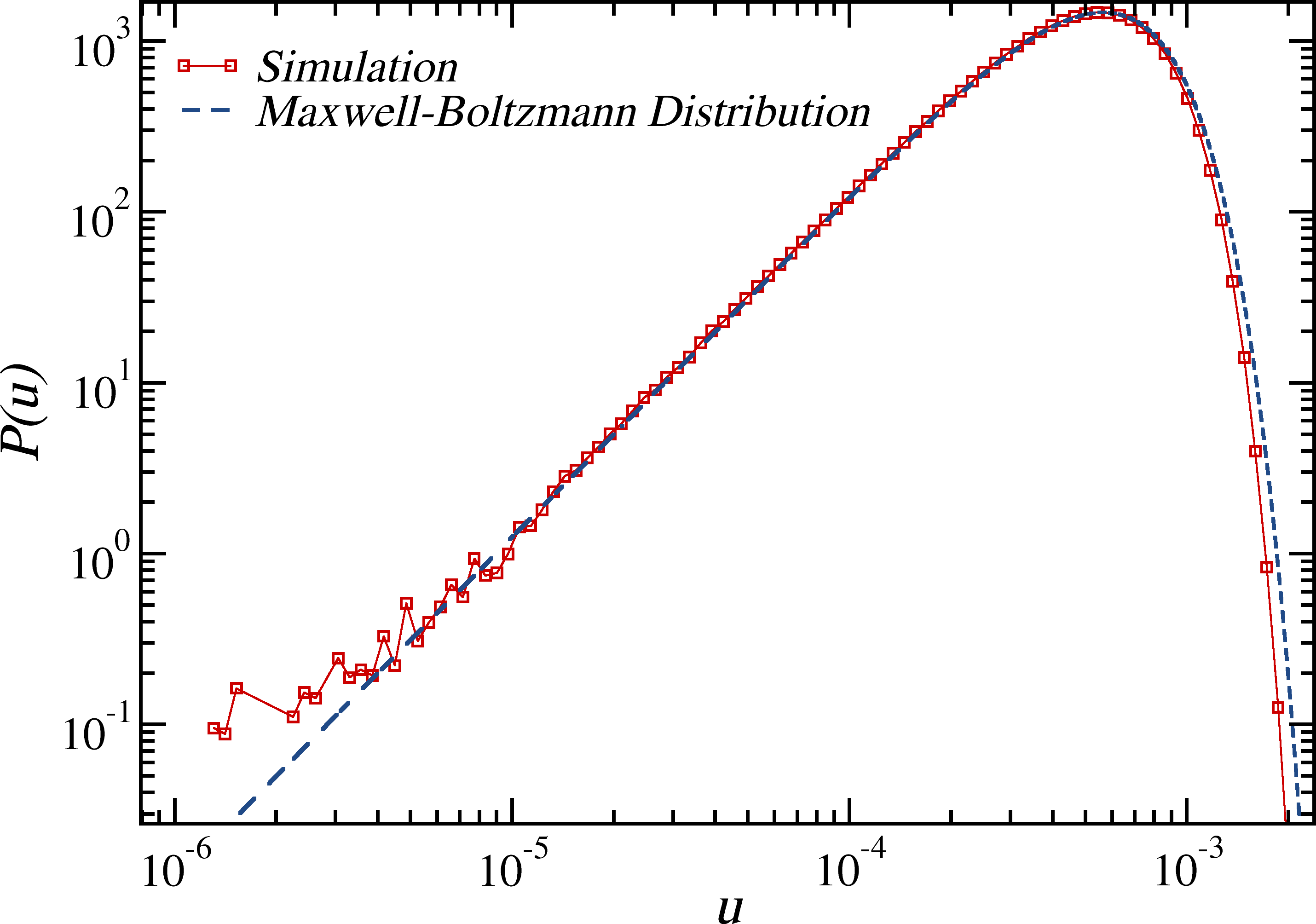}
\caption{\label{fig:pu_athermal_tracers} 
Distribution of (the absolute value of) tracers displacements per unit time. 
The red squares correspond to a simulation of a pure Brownian motion,
made with parameters $T= 8.0\times 10^{-7}$, $L = 32$. 
$u$ is defined as the displacement in $\Delta t=0.1$ units of time.
The blue dashed line corresponds to the Maxwell-Boltzmann 
distribution (Eq.~\ref{eq:maxwell_distribution}) with scale 
parameter c = 4.0$\times 10^{-4}$.
}
\end{figure}

The cases in which the movement of the tracer particles occurs only 
as a result of the thermal agitation (Brownian motion), display 
random trajectories. 
In that case, each displacement component ($u_x$, $u_y$, $u_z$) would 
populate a Gaussian distribution function.
Therefore, the module of the displacement $u= \left(u_x^2+u_y^2+u_z^2\right)^{1/2}$ must follow a Maxwell-Boltzmann distribution,
\begin{equation}
    P(u) = \sqrt{\frac{2}{\pi}} \frac{u^{2}}{c^{3}}\exp \left[ -\frac{u^{2}}{2c^{2}}\right ],
    \label{eq:maxwell_distribution}
\end{equation}
where $c$ is a scale parameter. 
Figure~\ref{fig:pu_athermal_tracers} shows a test simulation result for
$P(u)$ for tracer particles that undergo only thermal agitation. 
As expected, a good accuracy with respect to Eq.~\ref{eq:maxwell_distribution}
is seen. 

When, either because of high plastic activity or because of high thermal agitation,
the tracer's movement is random-walk-like we expect to observe a $P(u)$ following
Eq.~\ref{eq:maxwell_distribution}, at least in a the displacement $u$ range where 
the movement is effectively Brownian.

\section{Fully thermal system: variable activity}\label{app:ThermaCase_variable_activity}

\begin{figure}[t!]
\includegraphics[width=\columnwidth]{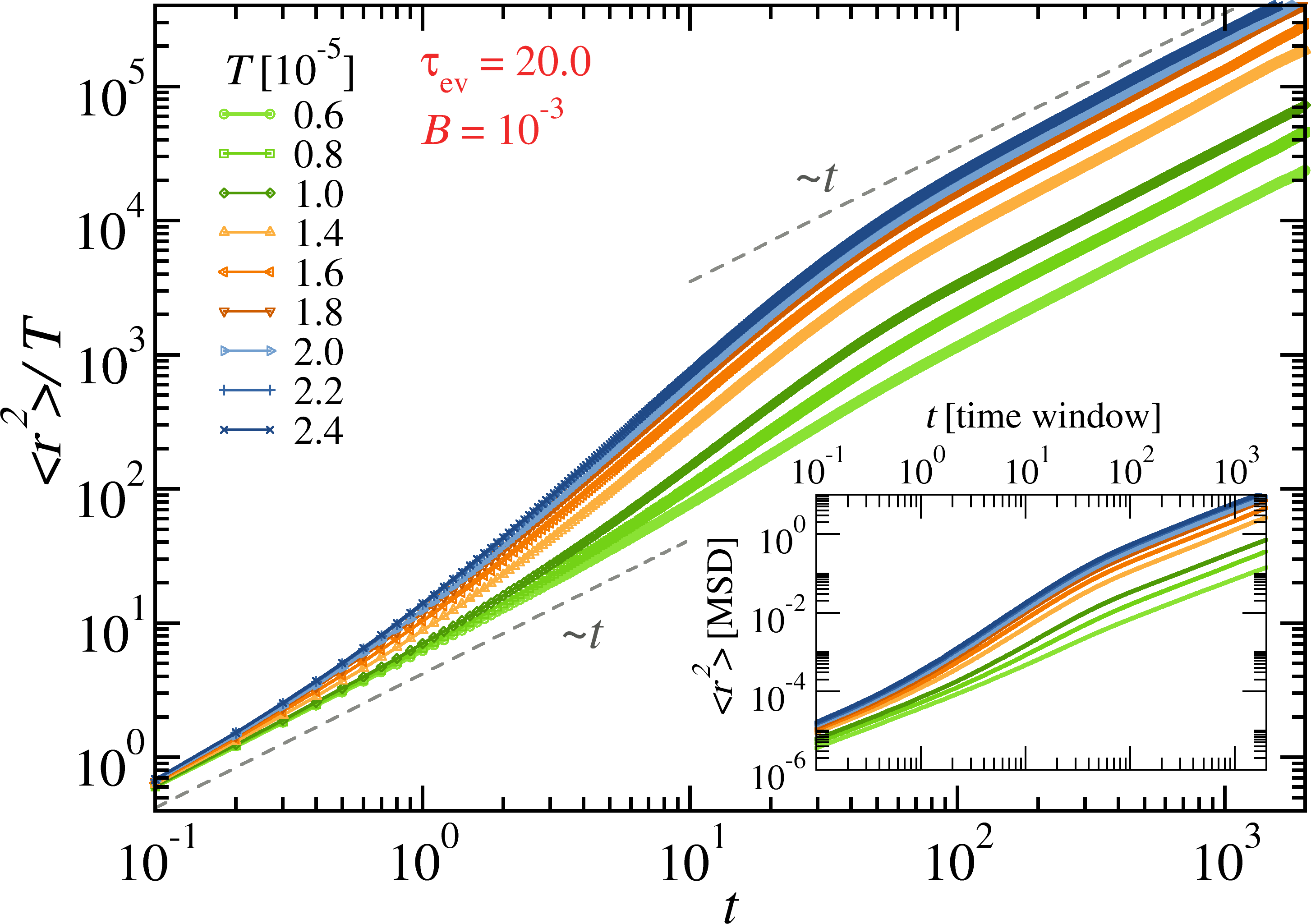}
\caption{\label{fig:msd-thermal-fixed-B} 
{\it Mean square displacement $\left< r^{2} \right>$ for a varying temperature and plastic activity.}
The main plot shows the mean square displacement normalized by temperature 
as a function of time observation window for different temperatures, ranging 
from $T = 6.0 \times 10^{-6}$ 
(light-green curve) to $T = 2.4 \times 10^{-5}$ (dark-blue curve).
The prefactor $B$ in Eq.~\ref{eq:activationrulesEPM} is fixed at $B=0.001$.
The gray dashed-lines are guidelines to show the ($\sim t$) behavior. 
The inset shows $\left<r^2\right>$ unscaled. 
 $\tau_{\tt ev} = 20.0$. $\epsilon_{0} = 0.1$. $L = 32$.
}
\end{figure}

The idea of a mean plastic activity that is insensitive to the 
external temperature is justified in cases where pre-stresses 
play a major role (e.g., in `as-quenched' glasses).
Nevertheless, one cannot rule out the case in which the same
temperature is controlling the thermal agitation of particles and 
the plastic activity.

In this work we have considered a thermal agitation for tracers implemented 
as a Brownian dynamics (fully overdamped).
Perhaps a more realistic approach would have been to use a  
Langevin dynamics (not necessarily fully overdamped). 
The control of the drag term might allow then to 
somehow decouple the ``thermal agitation'' 
temperature from the temperature parameter in the EP model.
Although we haven't explicitly used a new parameter for the drag term of 
the Brownian dynamics, the thermal agitation of the tracers and the 
temperature are relativized by the parameter ``B'' in the Arrhenius activation. 
\\

\begin{figure}[t!]
\includegraphics[width=0.39\textwidth]{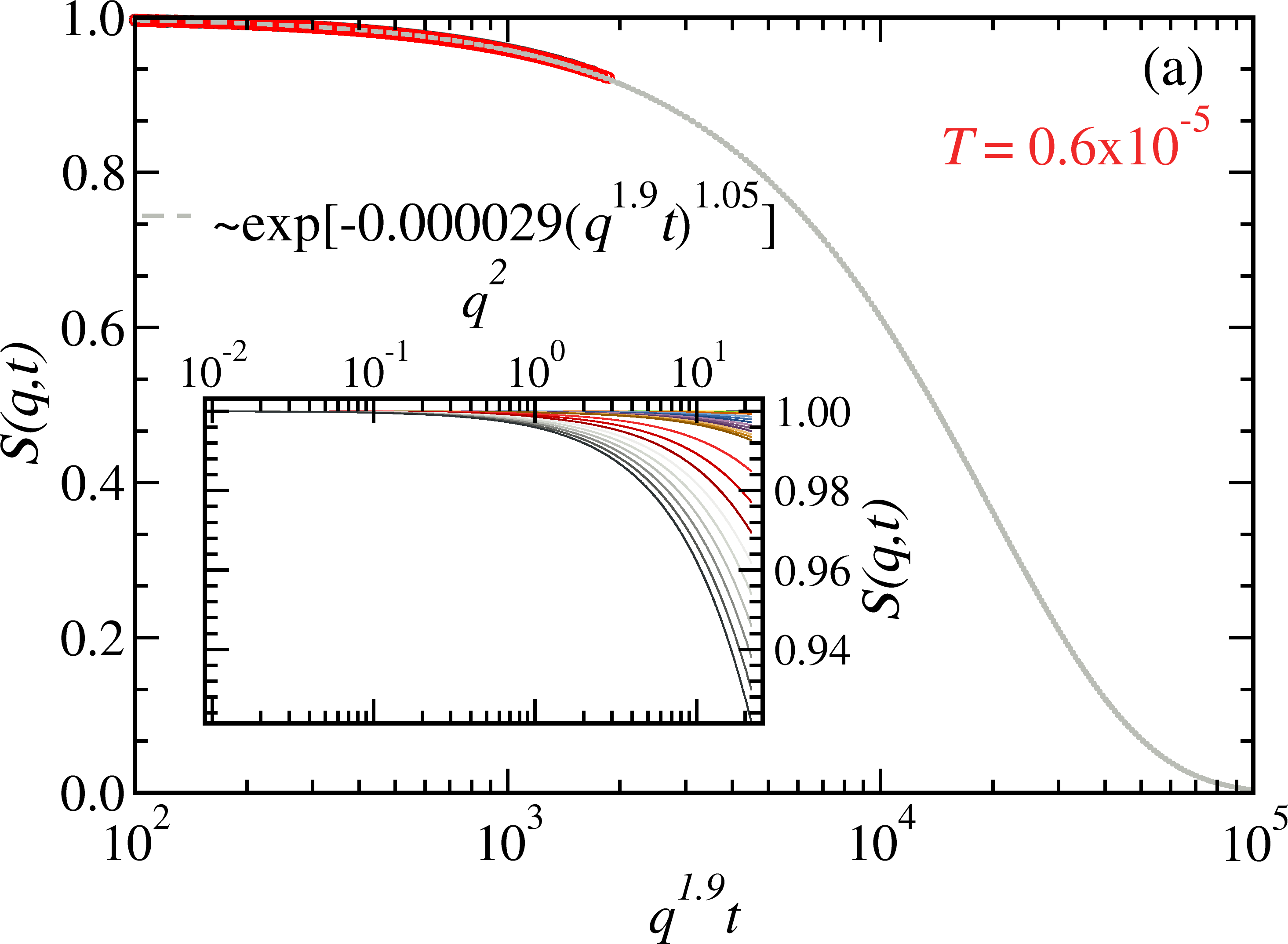}
\includegraphics[width=0.39\textwidth]{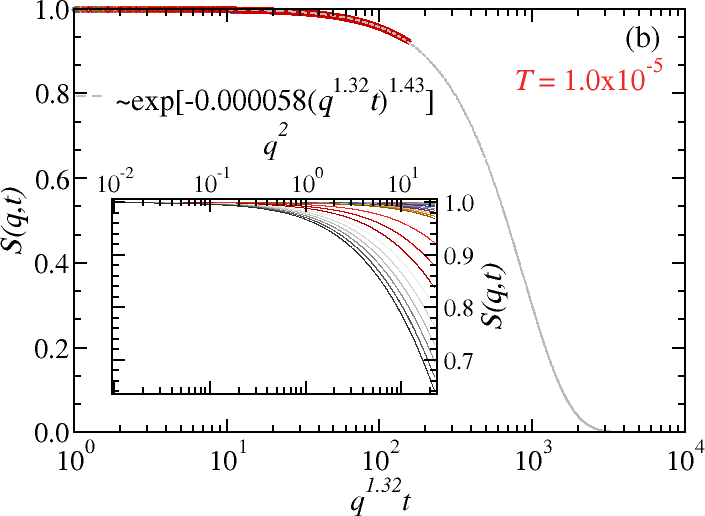}
\includegraphics[width=0.39\textwidth]{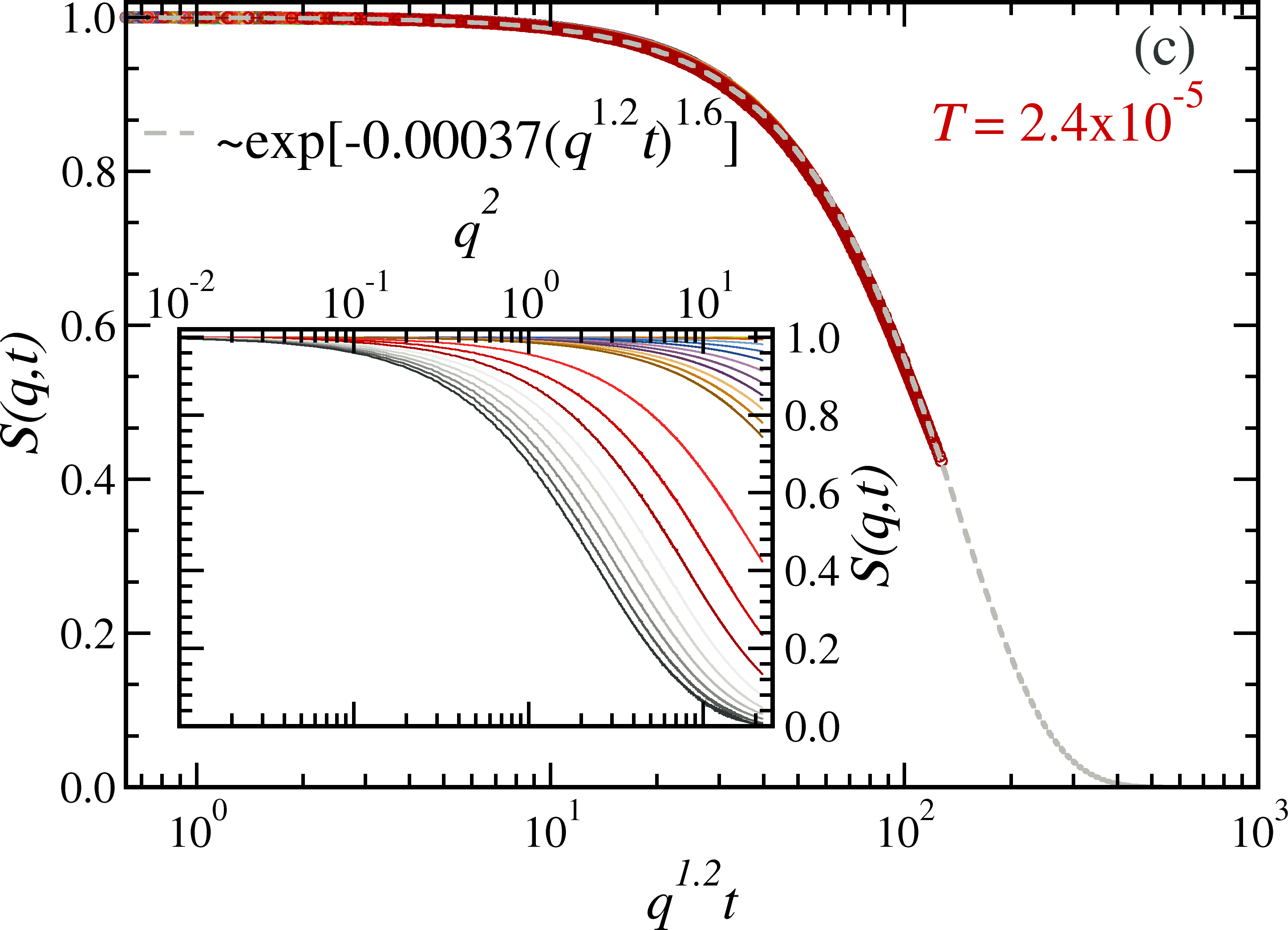}
\caption{\label{fig:Sq_T2.4e-5B1e-3_te20.0} 
{\it Dynamical structure factor $S(q,t)$ relaxation at short times 
and varying temperatures and activities}.
In all panels, the insets shows $S(q,t)$ as a function of $q^2$ for time 
windows up to $t=100$.
The dark-red curve corresponds to duration of plastic events, $\tau_{\tt ev} = 20$. 
$\epsilon_{0}$ = 0.1. $L = 32$.
}
\end{figure}

{\it Mean square displacement varying activity.--}
In Fig.~\ref{fig:msd-thermal-fixed-B} we show the mean square displacement 
for different temperatures, where each temperature controls both the thermal
agitation of the tracer particles and the probability of plastic events
by thermal activation (we have used now a fixed $B=1\times 10^{-3}$ in 
Eq.~\ref{eq:activationrulesEPM}).
Now the influence of both sources of tracer particles agitation, plastic activity
and temperature, is evidenced.
In fact, one can better distinguish the crossover between two diffusive regimes
previously insinuated in Fig.~\ref{fig:msd-thermal-fixed-ratio}.
At very low temperatures, plastic activity is almost absent or very spread and
$\left< r^{2} \right>$ at short times becomes essentially dominated by the
thermal agitation alone:
particles diffuse with a diffusion coefficient $D\propto T$ (which is shown
in the collapse of $\left< r^{2} \right>/T$ curves at short 
times~\footnote{Note that the use of an inertial dynamics for the tracers (instead of a
fully overdamped one) would change the behavior of the curves at short times.}
At higher temperatures, plastic activity becomes more and more important
and eventually dominates the diffusion at large enough time windows.
As a matter of fact, a $\left< r^{2} \right>/\left<a\right>$ scaling 
collapses the curves of higher temperatures at long times (not shown).
In between the two diffusive regimes, we can notice a super-diffusive crossover.
And even though it's not easy to identify a ballistic regime, that super-diffusivity
will already influence the relaxation shape exponents $\alpha$ and $\beta$.
Finally, the crossover between the activity-dominated to the temperature-dominated
diffusive regimes as we lower the temperature is understood by recalling that
plastic activity decays much faster than temperature 
($\left<a\right>\sim \exp(-1/T)$).
\\

{\it Dynamical structure factor varying activity.--}
Concerning the dynamical structure factor, 
as shown in Fig.~\ref{fig:Sq_T2.4e-5B1e-3_te20.0},
we now can see at short/intermediate times a compressed exponential 
behavior that turns to pure exponential as we decrease temperature. 
At contrast with the case in which the activity level is granted
on non-thermal grounds, here the temperature decrease simply suppresses
the plastic activity very fast and the few remaining persistent 
displacements of tracers easily disappear in a weak thermal agitation.

In contrast with what is observed in Fig.~\ref{fig:relaxation_exponent}, 
now the decrease in temperature implies an increase in $\alpha$ towards 
$\alpha \simeq 2$, and, consistently, $\beta$ approaches 1 as we lower $T$. 
Even with a much weaker diffusive coefficient and therefore producing a
much modest displacement and limited structure factor decay, 
for the lower temperature we obtain a nearly pure diffusive regime
at short times (Fig.~\ref{fig:Sq_T2.4e-5B1e-3_te20.0}a): 
$S(q,t) \propto \exp\left[-A\left(q^{2}t\right)\right]$.
One might say that at such low temperatures the tracers remain
near their initial position doing a small local Brownian motion. 
They will eventually diffuse further away, but in our observation
time window here, for the lowest temperatures they have barely 
traveled a distance comparable to the lattice cell. 
On the other hand, the compressed exponential is granted in the regime
in which plastic activity dominates the relaxation. 
For the highest temperature (Fig.~\ref{fig:Sq_T2.4e-5B1e-3_te20.0}c) 
the structure factor decays as $S(q,t) \propto \exp\left[-A\left (q^{1.2}t\right)^{1.6}\right]$.

%

\end{document}